\documentclass[traditabstract]{aa}
\usepackage{natbib}
\usepackage{graphicx}
\usepackage{txfonts}
\usepackage{mathrsfs}
\usepackage{mathbbol}
\DeclareMathAlphabet{\mathpzc}{OT1}{pzc}{m}{it}
\begin{document}
%
%________________________________________________________________
\def\etal{et al.}

\def\Mgv{M_{\rm g,500}}
\def\Mg{M_{\rm g}}
\def\YX {Y_{\rm X}}
\def\LX {L_{\rm X}}
\def\YSZ {Y_{\rm SZ}}
\def\TX {T_{\rm X}}
\def\fgv {f_{\rm g,500}}
\def\fg  {f_{\rm g}}
\def\kT {{\rm k}T}
\def\Mv {M_{\rm 500}}
\def \Rv {R_{500}}
\def \Rvs {R_{500}^{\rm Sim}}
\def \Rvh {R_{500}^{\rm HSE}}
\def \Mvs {M_{500}^{\rm Sim}}
\def \Mvh {M_{500}^{\rm HSE}}
\def\keV {\rm keV}
\def\ne {n_{\rm e}}
\def\rs {r_{\rm s}}

\def\MT {$M_{500}$--$T_{\rm X}$}
\def\MY {$M_{500}$--$Y_{\rm X}$}
\def\YM {$Y_{\rm SZ}$--$M$}
\def\YLX {$Y_{\rm SZ}$--$\LX$}
\def\MMg {$M_{500}$--$M_{\rm g,500}$}
\def\MgT {$M_{\rm g,500}$--$T_{\rm X}$}
\def\MgY {$M_{\rm g,500}$--$Y_{\rm X}$}
\def\LXM  {$\LX$--$\Mv$}
\def\msol {{\rm M_{\odot}}}

\def\lesssim{\mathrel{\hbox{\rlap{\hbox{\lower4pt\hbox{$\sim$}}}\hbox{$<$}}}}
\def\gtrsim{\mathrel{\hbox{\rlap{\hbox{\lower4pt\hbox{$\sim$}}}\hbox{$>$}}}}

% satellites
\newfont{\gwpfont}{cmssq8 scaled 1000}
\newcommand{\rexcess}{{\gwpfont REXCESS}}
\def \xmm {\hbox{\it XMM-Newton}}
\def \chandra {\hbox{\it Chandra}}
\def \rosat {\hbox{\it ROSAT}}
\def \planck {\hbox{\it Planck}}
\newcommand{\reflex}{{\gwpfont REFLEX}}

%-----------------------------------------------------------------------------------------------------------------------------------------------------------------------------------
%-----------------------------------------------------------------------------------------------------------------------------------------------------------------------------------

\title{The universal galaxy cluster pressure profile from a representative sample of nearby systems (REXCESS) and the $Y_{\rm SZ}$--$M_{500}$ relation } 
\author{M. Arnaud \inst{1},  G.W. Pratt \inst{1,2}, R. Piffaretti \inst{1},  H. B\"ohringer \inst{2}, J.H. Croston \inst{3}
  and   E. Pointecouteau \inst{4}       }
\offprints{M. Arnaud, \email{Monique.Arnaud@cea.fr}}
\authorrunning{M. Arnaud et al.}
\titlerunning{Pressure properties of the \rexcess}

 \institute{
 $^1$ Laboratoire AIM, IRFU/Service d'Astrophysique - CEA/DSM - CNRS
 - Universit\'{e} Paris Diderot, B\^{a}t. 709, CEA-Saclay, F-91191
 Gif-sur- Yvette Cedex, France \\ 
  $^2$ Max-Planck-Institut f\"ur extraterrestriche Physik, Giessenbachstra{\ss}e, 85748 Garching, Germany\\
$^3$ School of Physics and Astronomy, University of Southampton, Southampton, SO17 1BJ, UK\\
 $^4$ Universit\'{e} de Toulouse, CNRS, CESR, 9 av. du colonel Roche, BP 44346, 31028 Toulouse Cedex 04, France} 

  \date{Received 7 October 2009; accepted XX  2009}
  \abstract 
 {We investigate the regularity of cluster pressure profiles with  \rexcess, a representative sample of 33 local ($z < 0.2$) clusters drawn from the \reflex\ catalogue and observed with \xmm. The sample spans a mass range of $10^{14}\msol<\Mv<10^{15}\msol$, where $\Mv$ is the mass corresponding to a density contrast of 500. We derive an average profile from observations scaled by mass and redshift according to the standard self-similar model, and find that the dispersion about the mean is remarkably low, at less than 30 per cent beyond $0.2\,\Rv$, but increases towards the centre. Deviations about the mean are related to both the mass and the thermo-dynamical state of the cluster.  Morphologically disturbed systems have systematically shallower profiles while cooling core systems are more concentrated.
The scaled profiles exhibit a residual mass dependence with a slope of $\sim 0.12$, consistent with that expected from the empirically-derived slope of the \MY\ relation; however, the departure from standard scaling decreases with radius and is consistent with zero at $R_{500}$.
The scatter in the core and departure from self-similar mass scaling is smaller compared to that of the entropy profiles, showing that the pressure is the quantity least affected by dynamical history and  non-gravitational physics. 
Comparison with scaled data from several state of the art numerical simulations shows good agreement outside the core. Combining the observational data in the radial range $[0.03$--$1]\,\Rv$ with simulation data in the radial range $[1$--$4]\,\Rv$, we derive a robust measure of the universal pressure profile, that, in an analytical form, defines the physical pressure profile of clusters as a function of mass and redshift up to the cluster 'boundary'. Using this profile and direct spherical integration of the observed pressure profiles,  we estimate the integrated Compton parameter $Y$ and investigate its scaling with $\Mv$ and $\LX$, the soft band X--ray luminosity. We consider both the spherically integrated quantity, $Y_{\rm sph}(R)$, proportional to the gas thermal energy, and the cylindrically integrated quantity, $Y_{\rm cyl}(R)=\YSZ D_{\rm A}^2$, which is directly related to the Sunyaev-Zel'dovich (SZ) effect signal. From the low scatter of the observed $Y_{\rm sph}(\Rv)$--$\YX$ relation we show that  variations in pressure profile shape do not introduce extra scatter into the  $Y_{\rm sph}(\Rv)$--$\Mv$ relation as compared to that from the $\YX$--$\Mv$ relation.  The $Y_{\rm sph}(\Rv)$--$\Mv$ and $Y_{\rm sph}(\Rv)$--$\LX$ relations derived from the data are in excellent agreement with those expected from the universal profile. This profile is used to derive the expected $\YSZ$--$\Mv$  and \YLX\  relations for any aperture. 
}  

   \maketitle
%-----------------------------------------------------------------------------------------------------------------------------------------------------------------------------------
%-----------------------------------------------------------------------------------------------------------------------------------------------------------------------------------

\section{Introduction}

Galaxy clusters provide valuable information on cosmology, from the nature of dark energy to the physics driving galaxy and structure formation. Clusters are filled with a hot ionised gas that can be studied both in X-ray and through the thermal Sunyaev-Zel'dovich  (SZ) effect, a spectral distortion of the cosmic microwave background (CMB) generated via inverse Compton scattering of CMB photons by the free electrons. Its magnitude is proportional to the Compton parameter $y$, a measure of  the gas pressure integrated along the line-of-sight, $y = (\sigma_{\rm T}/m_{\rm e} c^2)\int P dl$, where $\sigma_{\rm T}$ is the Thomson cross-section, $c$ the speed of light,  $m_{\rm e}$ the electron rest mass and $P=\ne T$ is the product of the electron number density and temperature. The total SZ signal, integrated over the cluster extent, is proportional to the integrated Compton parameter $\YSZ$, $\YSZ D_{\rm A}^2 = (\sigma_{\rm T}/m_{\rm e} c^2)\int P dV$, where $D_{\rm A}$ is the angular distance to the system. 

As the gas pressure is  directly related to the depth of the gravitational potential,  $\YSZ D_{\rm A}^2$  is expected to be closely related to the mass. Numerical simulations \citep[e.g.,][]{das04, mot05, nag06, bon07} and analytical models \citep{rei06} of cluster formation indicate that the intrinsic scatter of the \YM\ relation is low, regardless of the cluster dynamical state \citep[see also][]{wik08} or the exact details of the gas physics. However, the normalisation of the relation {\it does} depend on the gas physics \citep{nag06,bon07}, as does the exact amount of scatter, the details of which are still under debate  \citep{sha08}. Given that this relation, and the underlying pressure profile, are key ingredients for the use of on-going or future SZ cluster surveys for cosmology, and provide invaluable information on the physics of the intra-cluster medium (ICM), it is important to calibrate these quantities from observations. 

In recent years, SZ observational capability has made spectacular progress, from the first spatially resolved  (single--dish) observations of individual objects \citep[][]{poi99,poi01,kom99, kom01} to the first discovery of new clusters with a blind SZ survey \citep[][]{sta09}. 
 Spatially resolved SZE observations directly probe  the mass weighted temperature along the line of sight. By contrast,  temperatures derived from X-ray spectra, by fitting an isothermal model to a multi-temperature plasma emission along the line of sight, are likely to be biased \citep{mat01}. Although schemes to correct for this effect have been defined  \citep{maz04,vik06b}, it remains a potential source  of systematics. 
 
Stacking analysis of WMAP data  around known X--ray clusters has allowed statistical detection of a scaled pressure profile \citep{afs07} or a spatially resolved decrement \citep{lie06,atr08,die09}, showing clear discrepancies with the prediction of a simple isothermal $\beta$--model.  Pressure or temperature profiles of individual clusters have started to be derived from combined analysis of X-ray and SZE imaging data, using non-parametric deprojection methods \citep{nor09} or more realistic  models than the $\beta$-model \citep{kit04, mro09}.   Interestingly, the profiles are found to be consistent with profiles derived using X-ray spectroscopic data \citep[see also][]{jia08,hal09}. However, such studies are still restricted to a few test cases,  particularly  hot  clusters. 

The \YM\ relation has been recently derived by \citet{bon08},  an important step forward as compared to previous work  based  on central decrement measurements using heterogenous data sets \citep{mcc03, mor07}; however, quantities were estimated within  $R_{2500}\sim 0.44\Rv$ \footnote{Here and in the following, $M_{\delta}$ and $R_{\delta}$ are the total mass and radius corresponding to a density contrast, $\delta$, as compared to $\rho_{\rm c}(z)$,   the critical density of the universe at the
cluster redshift: $M_{\delta}= (4\pi/3) \delta  \rho_c(z)  R_{\delta}^3$. $\Mv$ corresponds roughly to the virialised portion of clusters, and is traditionally used to define the 'total' mass.} and assuming an isothermal $\beta$--model, which may provide a biased estimate \citep{hal07}. In addition, the first scaling relation using weak lensing masses, rather than X--ray hydrostatic masses, has now appeared \citep{mar09}, although constraints from these data are currently weak. \\

In this context, statistically more precise, albeit indirect, information can be obtained from X-ray observations.  A key physical parameter is $\YX$,  the X--ray analogue  of the integrated Compton parameter,  introduced by \citet{kra06}.  $\YX$ is defined as the product  of  $\Mgv$,  the gas mass within $\Rv$ and $\TX$, the spectroscopic temperature outside the core. The local \MY\ relation for relaxed clusters has recently been calibrated \citep{nag07,mau07,app07, vik09}, with excellent agreement achieved between various observations  \citep[e.g., see][]{app07}.  However, the link between $\YX$ and $\YSZ$ depends on cluster structure through 
\begin{equation}
\frac{\YSZ D_{\rm A}^2 }{\YX} =  \frac{\sigma_{\rm T}}{m_{\rm e} c^2}\frac{1}{ \mu_{\rm e} m_{\rm p}} \frac{\langle \ne T\rangle} {\langle\ne\rangle_{\Rv} \TX}
\label{eq:yszyx}
\end{equation}
\noindent where the angle brackets denote volume averaged quantities. 
From Eq.~\ref{eq:yszyx}, it is clear that an understanding of the radial pressure distribution and its scaling is important not only as a probe of the ICM physics, but also for exploitation of these data. High resolution measurements of the radial density and temperature distribution are now routinely available from X--ray observations  but the pressure profile structure and scaling have been relatively little studied. The pressure profiles of groups have been studied by \citet{fin06} and \citet{joh09}. In the cluster regime,  \citet{fin05} analysed the 2D  pressure distribution in a flux-limited sample of 6 hot ($\kT>7 \keV$) clusters at $z\sim0.3$ showing  fluctuations at the $30\%$ level around the mean profile, scaled by temperature.  To our knowledge, the only study of pressure profiles scaled by mass is that of \citet{nag07}, who used {\it Chandra} X-ray observations to derive a universal pressure profile, with the external slope derived from numerical simulations. However, their sample was restricted to hot ($kT>5\keV$) relaxed clusters, which are all  cool  core systems, and contained five objects. For the reasons mentioned above, it is of considerable interest to extend this analysis to data from a larger and more representative sample of the cluster population.  

In this paper we do this by investigating the regularity of cluster pressure profiles with  \rexcess\  \citep{boh07}, a representative sample of 33 local ($z < 0.2$) clusters drawn from the \reflex\ catalogue \citep{boh04} and observed with \xmm. We derive an average profile from observations scaled by mass and redshift according to the self-similar model and  relate the deviations about the mean to both the mass and the thermo-dynamical state of the cluster (Sec.~\ref{sec:psc}). Comparison with data from several state of the art numerical simulations  (Sec.~\ref{sec:simul}) shows good agreement outside the central regions, which is the most relevant aspect for the $\YSZ$ estimate. Combining the observational data in the radial range $[0.3$--$1]\,\Rv$ with simulation data in the radial range $[1$--$4]\,\Rv$  allows us to derive a robust measure of the universal pressure profile up to the cluster 'boundary' (Sec.~\ref{sec:puniv}). Using this profile or direct spherical integration of the observed pressure profiles,  we estimate the spherically and cylindrically integrated Compton parameter and investigate its scaling with $\YX$, $\Mv$ and $\LX$, the soft band X--ray luminosity (Sec.\ref{sec:ysz}). 

We adopt a $\Lambda$CDM cosmology with $H_{0} = 70$~ km/s/Mpc, $\Omega_{\rm M} = 0.3$ and $\Omega_\Lambda= 0.7$. $h(z)$ is the ratio of the Hubble constant at redshift $z$ to its present value, $H_{0}$. $\TX$ is the temperature  measured in the $[0.15$--$0.75]~\Rv$ aperture.  All scaling relations are derived using the BCES orthogonal regression method with bootstrap resampling \citep{akr96}, and uncertainties are quoted throughout at the 68 per cent confidence level.

%-----------------------------------------------------------------------------------------------------------------------------------------------------------------------------------
%-----------------------------------------------------------------------------------------------------------------------------------------------------------------------------------
\section{The \rexcess\ data set}
\label{sec:data}

A description of the \rexcess\ sample, including \xmm\  observation
details, can be found in \citet{boh07}. The two clusters RXCJ0956.4-1004 (the Abell 901/902 supercluster) and J2157.4-0747 (a bimodal cluster)  are excluded from the present analysis. Cluster subsample classification follows the definitions described in \citet{pra09}: objects with centre shift parameter $\langle w\rangle > 0.01 \Rv$ are classified as morphologically disturbed, and those with central density $h(z)^{-2}\, n_{e,0} > 4 \times 10^{-2} {\rm cm^{-3}}$ as cool core systems. 

%================================================================================
%================================================================================
 \begin{figure}[t]
\centering
\includegraphics[ width=\columnwidth, keepaspectratio]{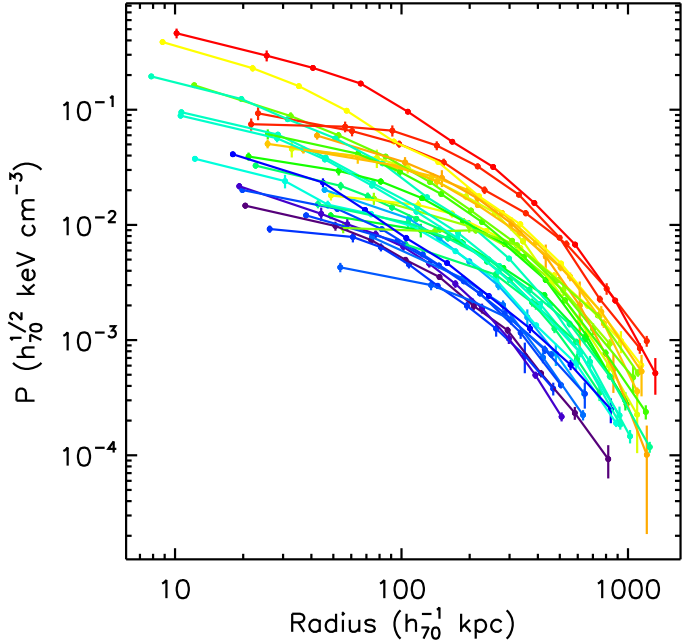} 
 \caption{\label{fig:pphys} \footnotesize The pressure profiles of the \rexcess\ sample. Pressures are estimated at the effective radii  of the temperature profile (points with errors bars). A line connects the data  points for each cluster to guide the eye. The data are colour coded (from cold--blue to hot--red) according to the spectroscopic temperature, $\TX$.} 
\end{figure}
%================================================================================
%================================================================================

The gas density profiles, $n_{\rm e}(r)$,  were derived by \citet{cro08} from  the  surface brightness profiles using the non-parametric deprojection and PSF-deconvolution technique introduced by  \citet{cro06}. The density at any radius of interest is estimated by interpolation in the log-log plane. The procedure to extract the  2D temperature profiles is detailed in \citet{pra09}. The  3D profiles, $T(r)$,  were derived by fitting  convolved parametric models  \citep{vik06a} to these data, taking into account projection and PSF effects  \citep{poi04} and weighting the contribution of temperature components to each ring as proposed by \citet{vik06b} to correct for the spectroscopic bias mentioned above.  A Monte Carlo procedure is used to compute the  errors, which are then corrected to take into account the fact that parametric models over-constrain the 3D profile.  Full details will be given in a forthcoming paper.  As the temperature profiles are measured on a lower resolution  radial grid than the density profiles,  the pressure profiles, $P(r)=n_{\rm e}(r)T(r)$,  are estimated at  the weighted effective radii \citep{lew03} of each annular bin of the 2D temperature profiles. They are   presented in Fig.~\ref{fig:pphys}.

Since the sample contains systems in a variety of dynamical states, we choose to use $\YX$ as a mass proxy rather than the hydrostatic mass. Extensive discussion of how this could affect our results is presented in Sec.~\ref{sec:pscmass}. For each cluster, $\Mv$ is estimated iteratively from the \MY\ relation, as described in \citet{kra06}.   We used the updated calibration of  the \MY\ relation, obtained by  combining the  \citet{app07} data on  nearby relaxed clusters observed with \xmm\ with new \rexcess\ data (Arnaud et al., in prep). The sample comprises 20 clusters:  8 clusters from \citet{app07}, excluding the two lowest mass clusters whose $\Mv$  estimate requires extrapolation, and the 12  relaxed \rexcess\ clusters with  mass profiles measured at least down to $\delta = 550$. The derived  \MY\  relation 

{\small
\begin{equation}
h(z)^{2/5}\,\Mv = 10^{14.567\pm0.010}\left[\frac{\YX}{2\times10^{14}\,{\rm h_{70}^{-5/2}}\,{\msol}\,\keV}\right]^{0.561\pm0.018}\! {\rm h_{70}^{-1}\,\msol}
\label{eq:mynstd}
\end{equation}
}

\noindent is consistent with the relation derived by \citet{app07} but with improved accuracy on slope and normalization.

The slope differs from that expected in the standard self-similar model ($\alpha=3/5$) by only $\sim 2 \sigma$. We will thus also consider the \MY\ relation obtained by fixing the slope to its standard value:

{\small
\begin{equation}
h(z)^{2/5}\,\Mv = 10^{14.561\pm0.009}\left[\frac{\YX}{2\times10^{14}\,{\rm h_{70}^{-5/2}}\,{\msol}\,\keV}\right]^{3/5}\,{\rm h_{70}^{-1}\,\msol}
\label{eq:mystd}
\end{equation}
}
%
 %================================================================================
%================================================================================
   \begin{figure}[t]
\centering
\includegraphics[ width=\columnwidth, keepaspectratio]{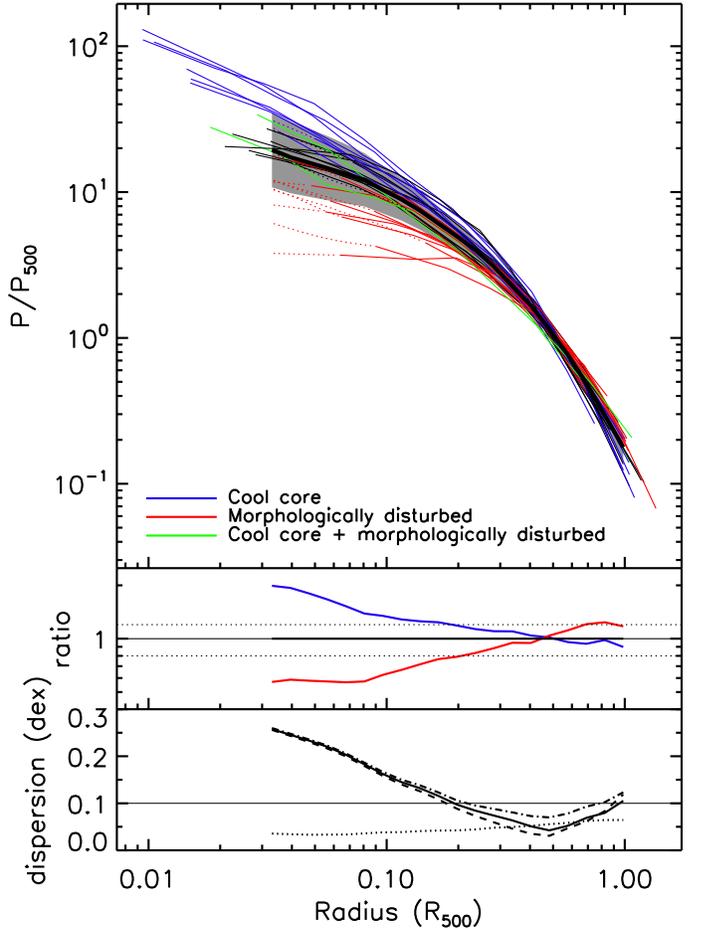} 
 \caption{\label{fig:pscdyn} \footnotesize The scaled pressure profiles of the \rexcess\ sample, colour coded according to the (thermo)dynamical state (see labels and Sec.~\ref{sec:data}). The radii are scaled to $\Rv$ and the pressure to $P_{500}$ as defined in Eq.~\ref{eq:p500}, with $\Mv$ estimated from the  \MY\ relation (Eq.~\ref{eq:mynstd}). Full lines: pressure profile as in Fig.~\ref{fig:pphys} with data points omitted for clarity. Dotted lines: extrapolated pressure (see text).
  The thick black line is the average scaled profile and the grey area corresponds to the $\pm1\sigma$ dispersion around it. Middle panel: ratio of the average profile of cool core (blue) and disturbed (red)  systems to the overall average profile. Bottom panel: The solid line is the statistical dispersion as a function of scaled radius. Dotted line:  additional dispersion expected from the intrinsic dispersion in the \MY\ relation. Dash-dotted line: quadratic sum of the two dispersions. Dashed line: dispersion obtained for $\Mv$ estimated from the  standard slope \MY\ relation (Eq.~\ref{eq:mystd}). } 
\end{figure}
%================================================================================
%================================================================================

%-----------------------------------------------------------------------------------------------------------------------------------------------------------------------------------
%-----------------------------------------------------------------------------------------------------------------------------------------------------------------------------------
\section{Scaled pressure profiles}
\label{sec:psc}
\subsection{Scaled profiles}

 %================================================================================
%================================================================================
\begin{figure*}[t]
\centering
\resizebox{\hsize}{!} {
\includegraphics{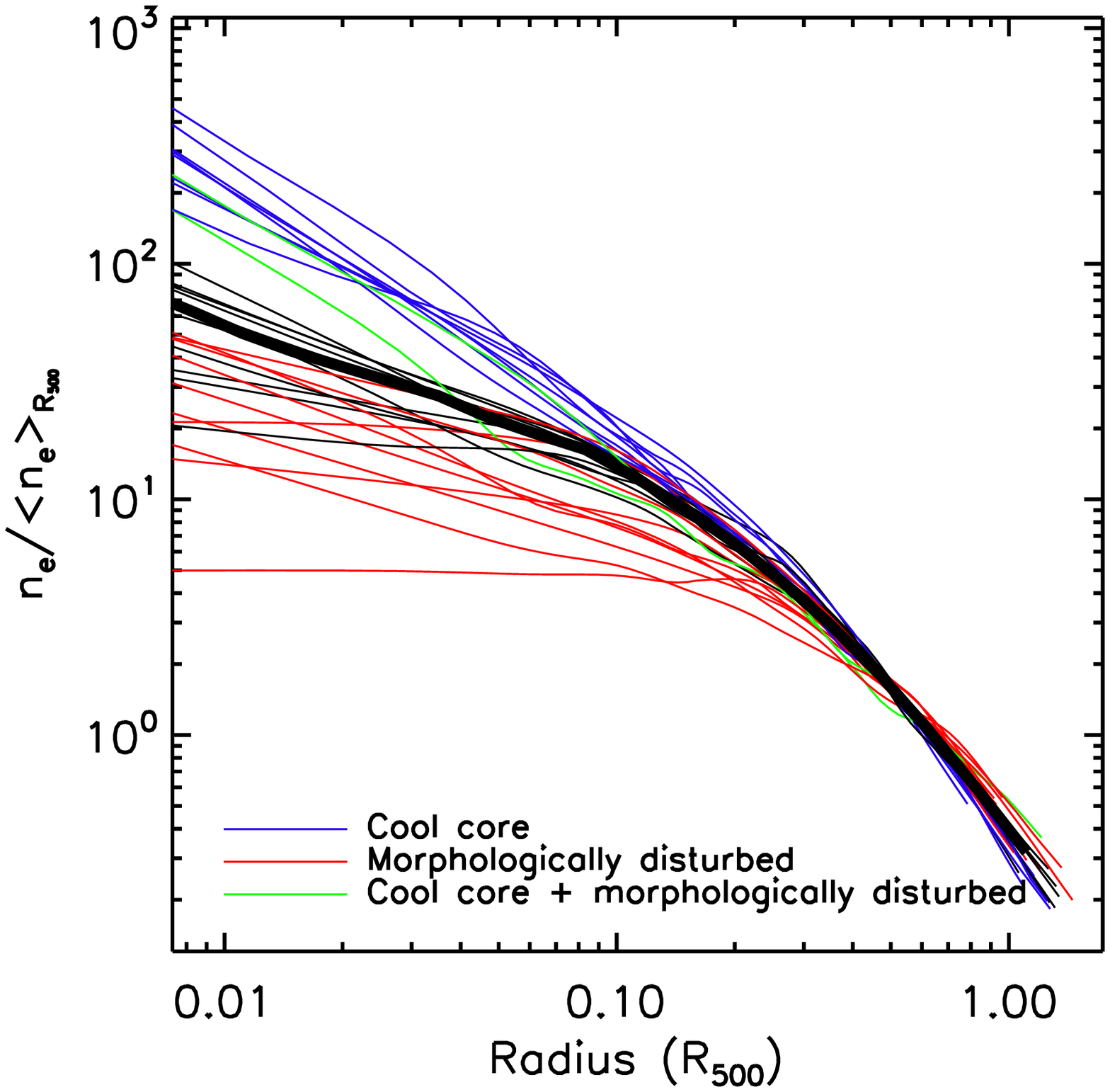} 
\hspace{8mm}
\includegraphics{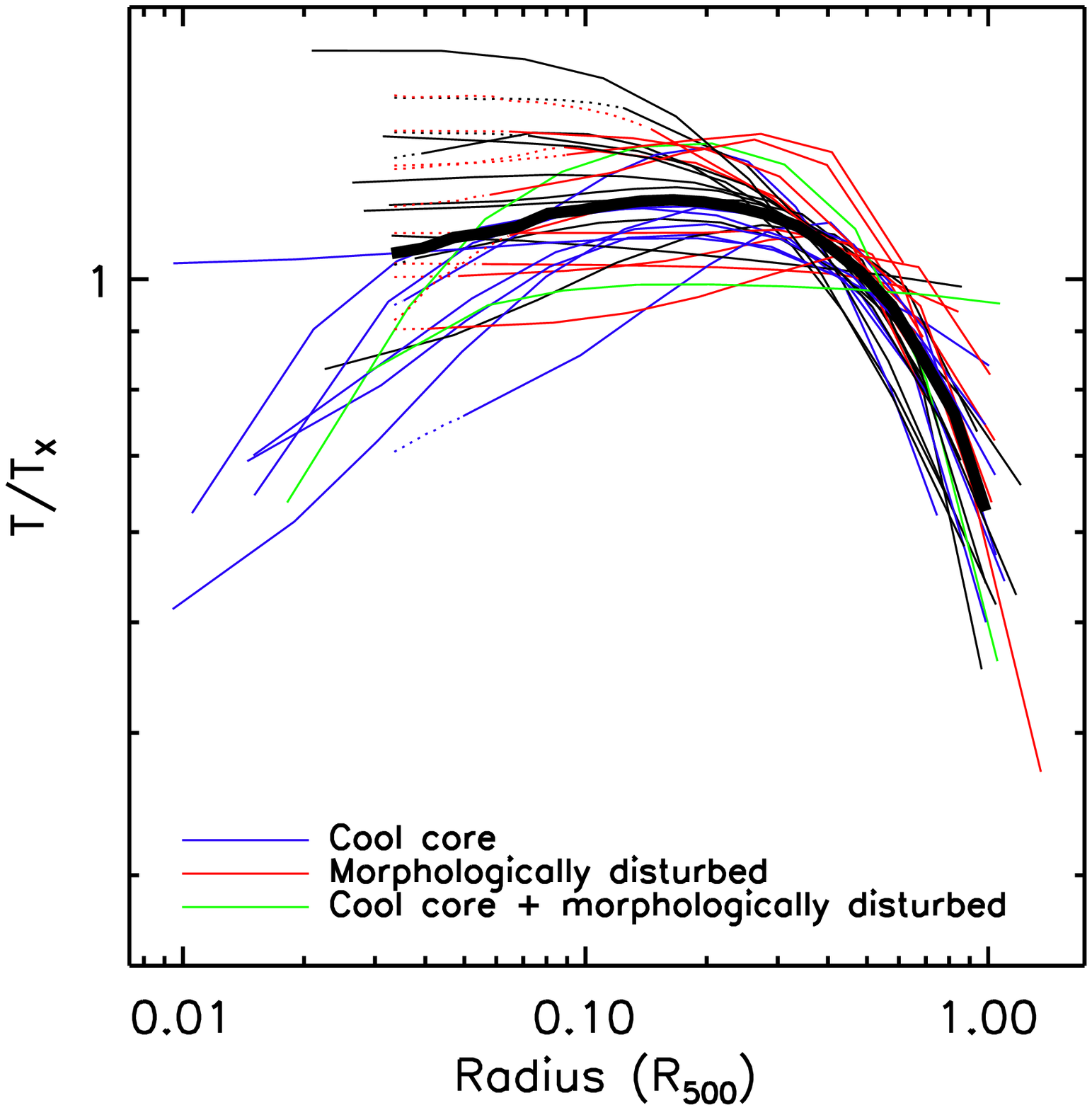} 
\hspace{8mm}
\includegraphics{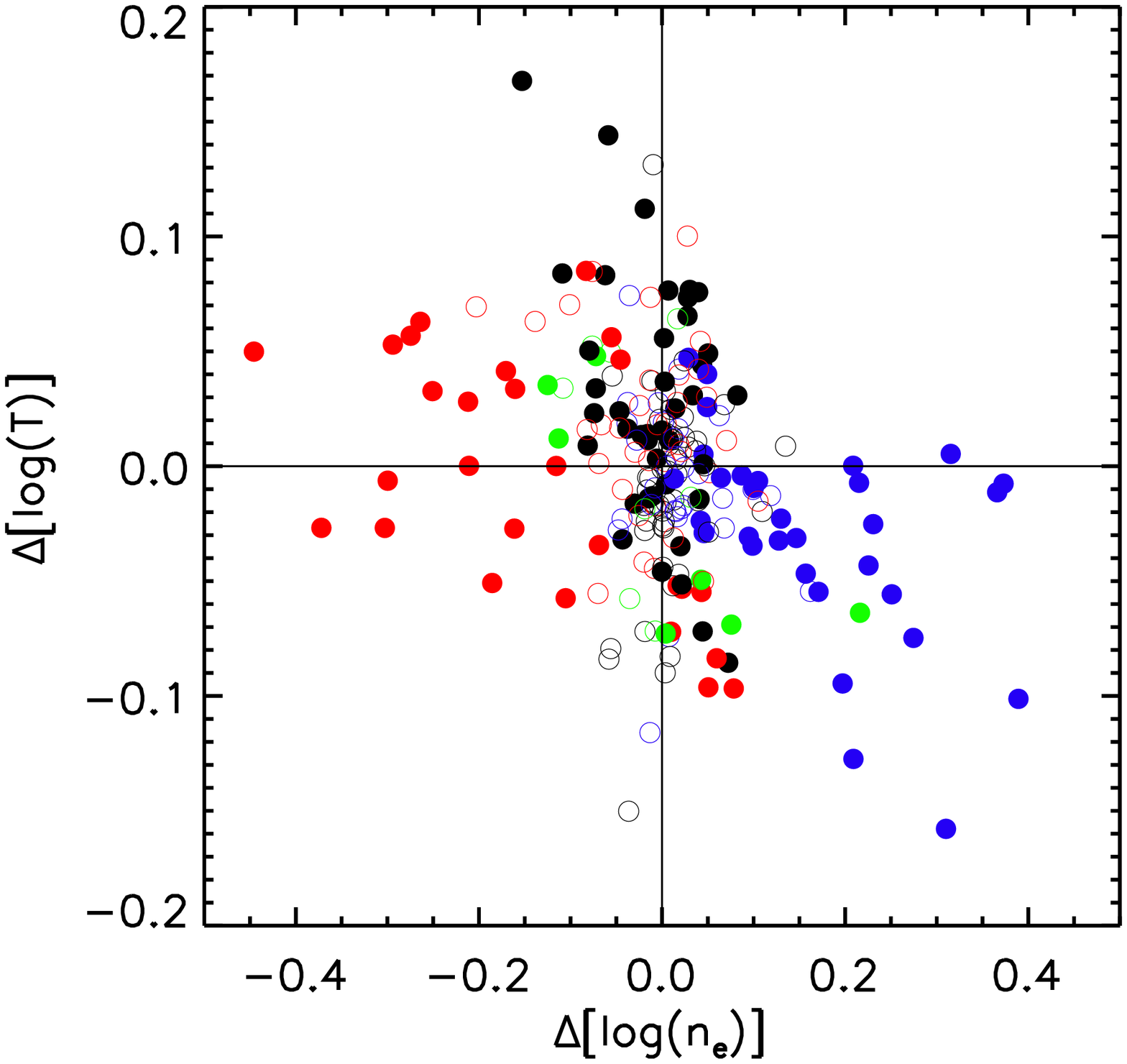} 
}
 \caption{\label{fig:ntdyn} \footnotesize The scaled density (left panel) and temperature (middle panel)  profiles of the \rexcess\ sample. Each profile is  colour coded according to the cluster (thermo)dynamical state (see labels and Sec.\ref{sec:data}). The radii are scaled to $\Rv$,  estimated from the \MY\ relation (Eq.~\ref{eq:mynstd}). The density is scaled to the mean density within $ \Rv$ and the temperature to $\TX$, the spectroscopic temperature measured in the $[0.15$--$0.75]~\Rv$ aperture. In each panel, the thick black line is the average scaled profile.  Right panel: logarithmic deviation of the scaled temperatures from the average scaled profile versus the corresponding deviation  for the density, at each effective radius of the temperature profile annular bins.  Data corresponding to $r/\Rv < 0.2$ and $r/\Rv> 0.2$ are marked with filled and open circles, respectively.  The deviations are anti-correlated in the core.} 
\end{figure*}
%================================================================================
%================================================================================

 The scaled pressure profiles
\begin{equation}
  p(x)= \frac{P(r)}{P_{500}}~~~{\rm where}~~x=\frac{r}{\Rv}
 \end{equation}
\noindent are presented in Fig.~\ref{fig:pscdyn}. The pressure is normalised  to the characteristic pressure $P_{500}$,  reflecting the mass variation expected in the standard self-similar model, purely based on gravitation \citep[][and Appendix~\ref{ap:ss}]{nag07}.
\begin{equation}
P_{500}  =  1.65\times10^{-3}\,h(z)^{8/3}\,\left[\frac{\Mv}{3\times10^{14}\,{\rm h^{-1}_{70}}\,\msol}\right]^{2/3}~~{\rm h_{70}^{2}\,\keV\,cm^{-3}}
\label{eq:p500}
\end{equation}
For comparison we also plot in Fig.~\ref{fig:ntdyn}  the scaled temperature profiles, $t(x)= T(r)/\TX$ as well as the scaled density profiles, $\widetilde{\ne}(x)$.  Note that the density profiles have been normalised to the mean density within $\Rv$, so that the dispersion is only due to  variations in shape \footnote{The normalisation of the density profiles, scaled according to the standard self-similar model, varies with mass as shown by \citet{cro08}.}.

The resolution in the centre and radial extent of the pressure profiles are determined by that of the temperature profiles, in practice the effective radius of the inner and outer annular temperature profile bins, which varies from cluster to cluster (see Fig.~\ref{fig:pscdyn}). In particular, the peaked emission of cool core clusters allows us to measure  the profiles deeper into the core than for disturbed clusters, which have more diffuse emission (see also Sec.~\ref{sec:pscdyn}). 

\subsection{Average scaled pressure profile}
\label{sec:pscm}

We computed an average scaled pressure profile,  $\mathbb{p}(x)$,  from the median value of the scaled pressure in  the radial range where data are available for at least 15 clusters without extrapolation (about $[0.03$--$1]~\Rv$).  However, to avoid a biased estimate of the average profile in the core, where the dispersion is large and more peaked clusters are measured to lower radii (Fig.~\ref{fig:pscdyn}), it is important to include all clusters in the computation. For this purpose, we extrapolated  the pressure profiles in the core using the best fitting temperature model used in the deprojection of the temperature profile. This extrapolation is only weakly model dependent since it essentially concerns  disturbed clusters (Fig.~\ref{fig:pscdyn}) which are observed to have rather flat central temperature profiles (Fig.~\ref{fig:ntdyn}). The average profile is plotted as a thick line in Fig.~\ref{fig:pscdyn}. The dispersion around it is defined as the plus or minus standard deviation from the average profile, computed in the log-log plane. 

%================================================================================
%================================================================================
 \begin{figure*}[tp]
\begin{center}
\begin{minipage}[t]{0.95\hsize}
%\begin{minipage}[t]{\hsize}
\resizebox{\hsize}{!} {
\includegraphics{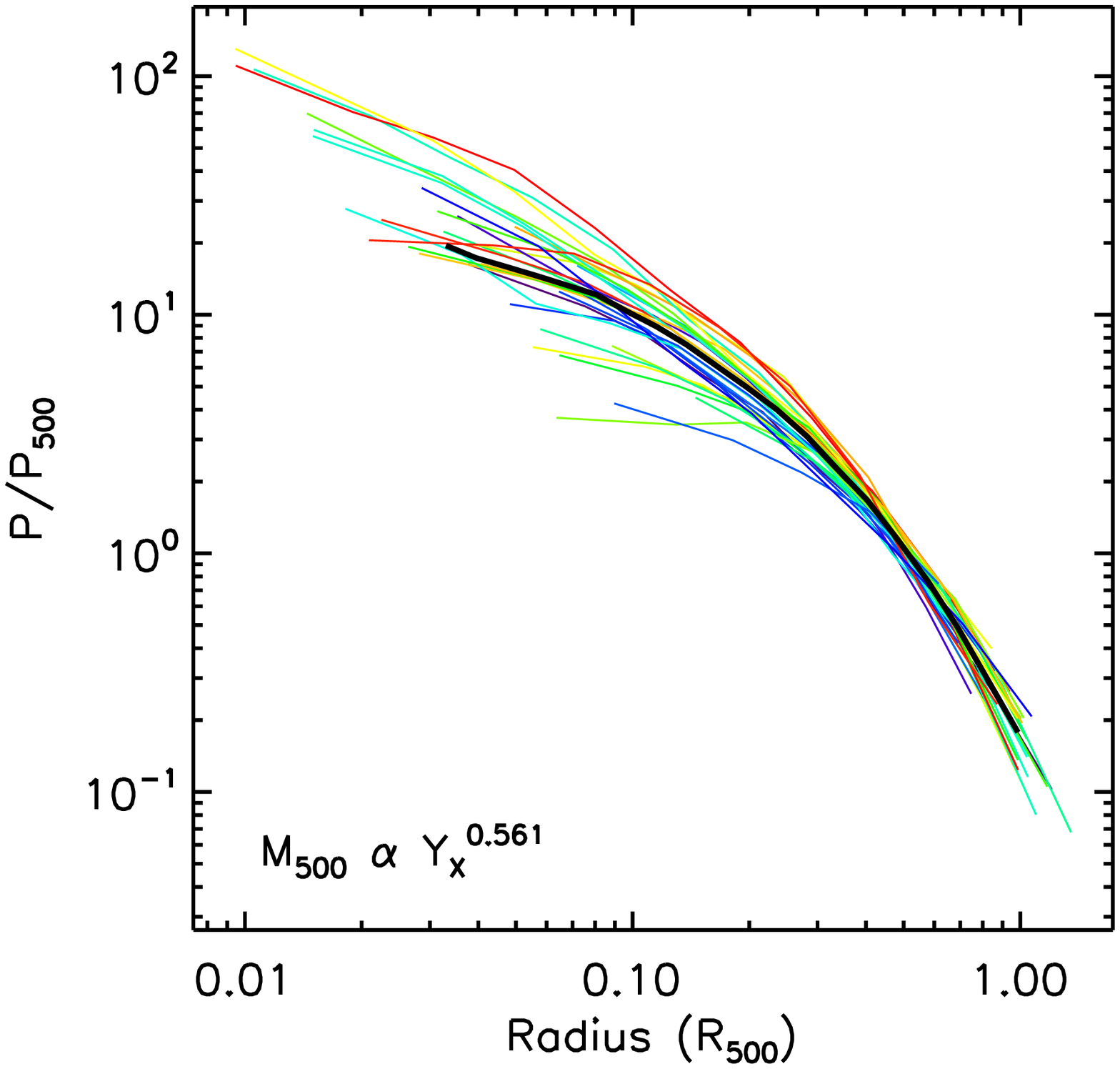} 
\hspace{8mm}
\includegraphics{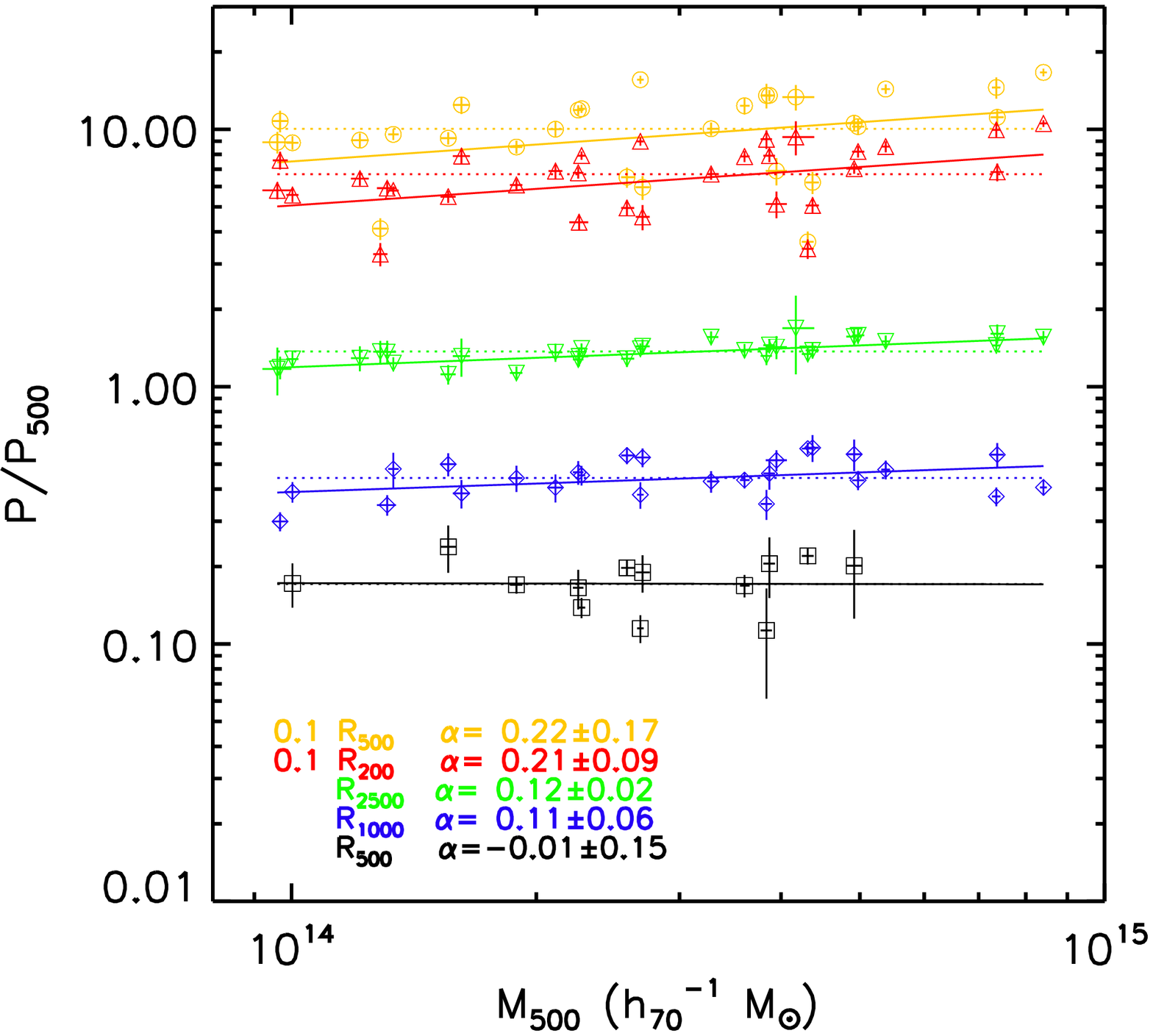}
} 
\end{minipage}\\[2mm]
\begin{minipage}[t]{0.95\hsize}
%\begin{minipage}[t]{\hsize}
\resizebox{\hsize}{!} {
\includegraphics{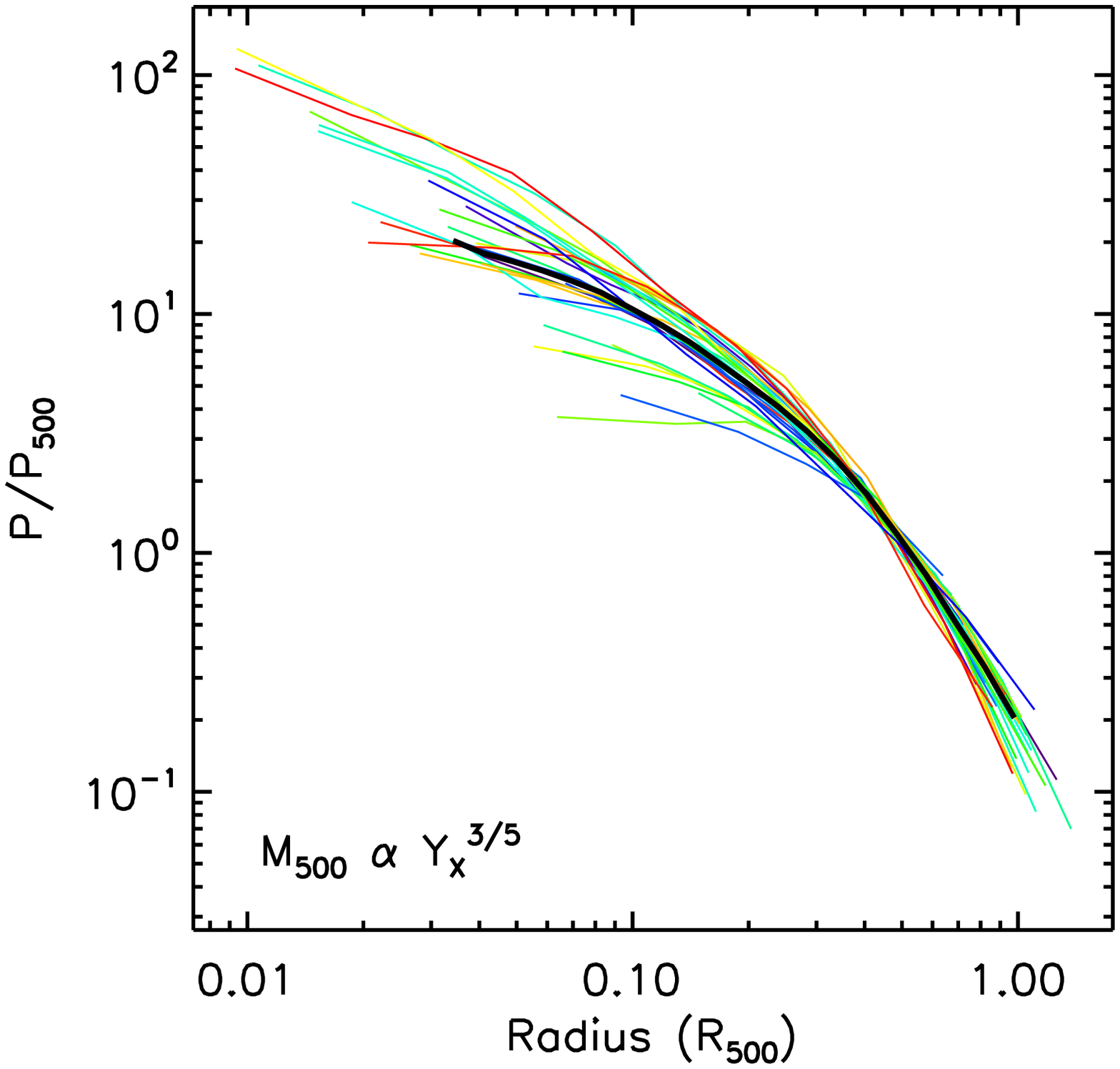} 
\hspace{8mm}
\includegraphics{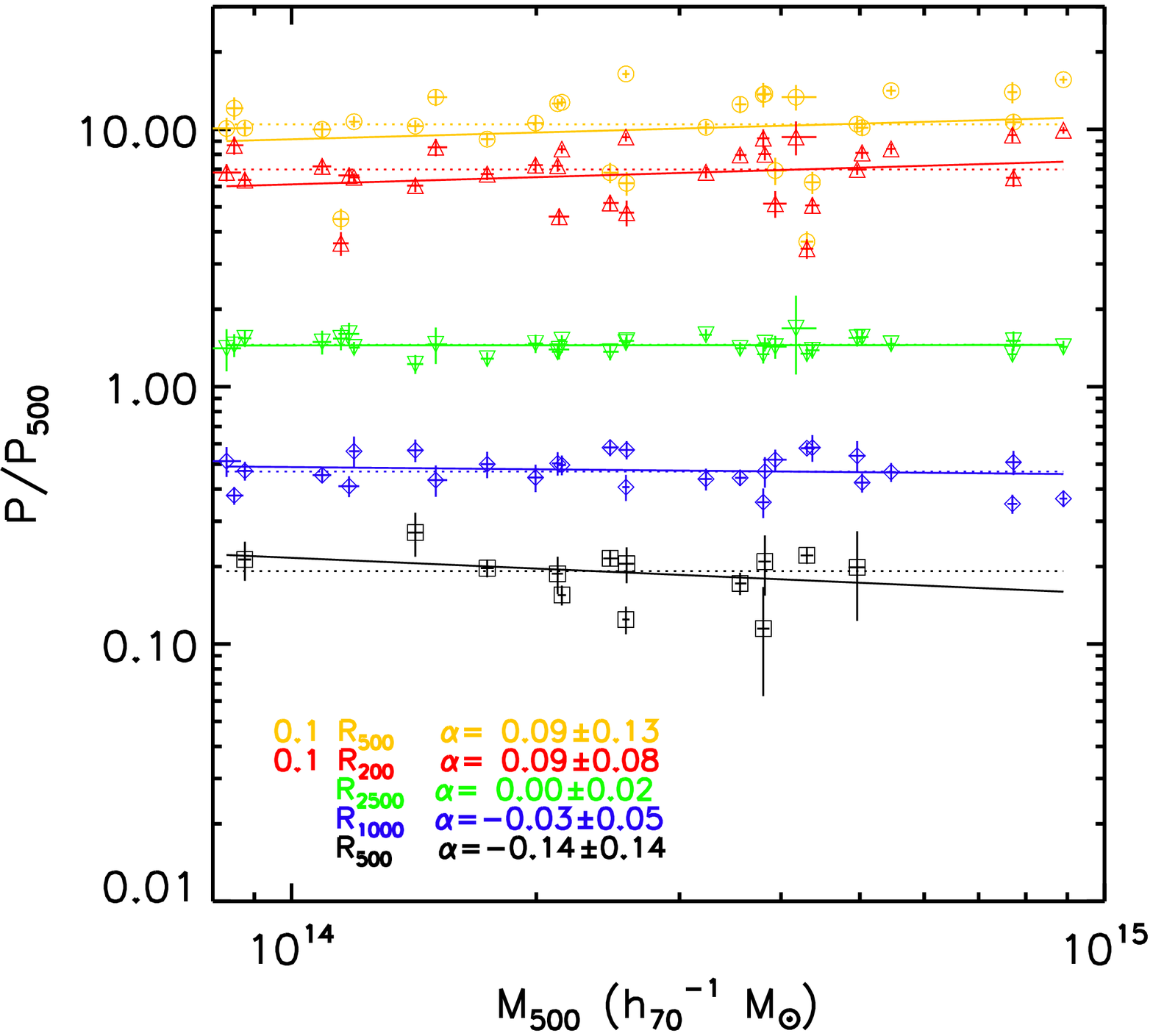}
} 
\end{minipage}
 \caption{\footnotesize The scaled pressure profiles  of the \rexcess\ sample, colour coded according to spectroscopic temperature measured in the $[0.15$--$0.75]$ $R_{500}$ aperture (left panels). Right panels: corresponding scaled pressure estimated at different values of scaled radii plotted as a function of cluster mass. Full lines:  power law fit at each scaled radius, with the best fitting slope given in the labels.  Dotted line: value for the average scaled profile at that radius. The cluster masses, $\Mv$,  are  estimated from the \MY\ relation,  either the best fitting empirical  relation  (Eq.~\ref{eq:mynstd}, top panels) or the relation obtained from fixing the slope to its standard value (Eq.~\ref{eq:mystd}, bottom panels).  } 
 \label{fig:pscmass}
\end{center}
\end{figure*}
 %================================================================================
%================================================================================

\subsection{Dispersion, radial structure and dynamical state}
\label{sec:pscdyn}

For a perfectly self-similar cluster population, the scaled profiles should coincide. The dispersion around the average scaled profile  is less than $25\%$ beyond the core ($r>0.2\Rv$) and increases towards the centre (bottom panel of Fig.~\ref{fig:pscdyn}). This dispersion reflects  a variation of shape with cluster (thermo)dynamical state, as clearly seen in  Fig.~\ref{fig:pscdyn}: shallower profiles, at all radii, are observed for morphologically  disturbed clusters while the cooling core clusters have the most concentrated profiles. The typical difference between the average profiles of these two populations is $\sim 20\%$ in the outskirts and as high as a factor of four at $0.03\,\Rv$ (Fig.~\ref{fig:pscdyn} middle panel). 

When compared to the density profiles (Fig.~\ref{fig:ntdyn}, left panel) the pressure profiles are distinctly more regular and present  less dispersion in the core.  The reason lies in the anti-correlation between the deviation of scaled  temperatures and densities from their respective average scaled profiles, $\mathbb{n}(x)$ and $\mathbb{t}(x)$, as shown Fig. ~\ref{fig:ntdyn} (right panel).  For data interior to $r<0.2\Rv$, a Spearman rank test finds a probability of   $10^{-7}$ that the anti-correlation between $\log(\widetilde{\ne}(x) - \log(\mathbb{n}(x))$ and $\log(t(x) - \log(\mathbb{t}(x))$ occurs by chance. The correlation disappears at large radii (probability of 0.6 for $r>0.2\Rv$).  Qualitatively, this is the result of the well-known fact that cool core clusters have peaked density profiles, with a temperature drop in the centre, while unrelaxed objects have flatter density cores (Fig.~\ref{fig:ntdyn}, left panel) and  constant or increasing temperature toward the centre  (Fig.~\ref{fig:ntdyn}, middle panel). 

\subsection{Dependence on mass and mass-proxy relation}
\label{sec:pscmass}

Since we derived $\Mv$ from the \MY\  relation,  the scaling quantities $\Rv$ and $P_{500}$ and the pressure profiles are not independent, as they are both related to the product of the gas density and temperature.  We first examine how this may affect our results. 
From the definition of the pressure $P(r)\!\!=\!\!\ne(r)T(r)$, and noting that $P_{500}\propto\Mv^{2/3}$ and that $\YX=\Mgv\TX\propto\langle\ne(r)\rangle_{\Rv}\Rv^3\TX\propto\langle\ne(r)\rangle_{\Rv}\TX\Mv$, where the angle brackets denote a volume average within $\Rv$, the scaled pressure $p(x) = P(x\Rv)/P_{500}$ is proportional to 
\begin{equation}
p (x) \propto \frac{P(x\Rv)}{\langle P(r)\rangle_{\Rv}}~\frac{\langle ne(r)T(r)\rangle_{\Rv}}{\langle \ne(r)\rangle_{\Rv} \TX}~\frac{\YX}{\Mv^{5/3}}.\label{eq:pm}
\end{equation}
This equation makes explicit the  link  between  the scaled pressure profiles and  the \MY\  relation. The first two dimensionless  terms in the right hand part of the equation purely depend on the internal gas structure within $\Rv$. They determine the average shape of the scaled profile. 
The third term depends on the global cluster scaling properties between $\YX\ $ and $\Mv$ and determine both the normalisation of the average scaled profile and the 'typical' mass dependence of the profiles (discussed  at the end of the section).  
 
Using $\Mv$ values derived from the \MY\ relation, rather than the 'true' $\Mv$  value, is equivalent to assuming a perfect correlation between $\Mv$ and $\YX$, i.e with no scatter. Provided that the correct \MY\  relation is used and that $\sigma_{\log,MY}$ does not depend on mass or dynamical state, use of the \MY\ relation will not introduce a systematic bias into the scaled profiles, but their dispersion will be underestimated. Let us define the intrinsic scatter of the \MY\ relation, $\sigma_{\log, MY}$,  as the standard deviation of $\log(\Mv)$ from the value from the best fitting relation at a given $\YX$.  We can estimate the additional dispersion due to $\sigma_{\log,MY}$  from the effect on the average scaled profile of a variation of $\log(\Mv)$ by  $\pm\sigma_{\log, MY}$. Since $\Rv\propto\Mv^{1/3}$ and $P_{500}\propto\Mv^{2/3}$, the  profile is translated in the log-log plane by $\pm1/3\sigma_{\log,MY}$ and $\pm2/3\sigma_{\log,MY}$ along the x and y axis, respectively.  Assuming $\sigma_{\log,MY}=0.04$ \citep[about $10\%$, ][]{kra06,app07}, the additional dispersion (in dex units), computed from the difference between the translated profiles at a given scaled radius, is plotted in the bottom panel of Fig.~\ref{fig:pscdyn}.  It is non-negligible beyond the core, but the total dispersion, estimated by summing quadratically this additional contribution, is expected to remain below $30\%$.  It is negligible in the core, where the dispersion  is dominated by structural variations. 

Finally, the \MY\  relation being  derived from mass estimated using the hydrostatic equilibrium, we expect an offset between that relation and the 'true' \MY\  relation. The $\Mv$ used in this study are thus likely to be underestimated.  The effect of such a bias is  to translate all the scaled profiles together (provided that it is a simple factor independent of mass). This will not affect any shape or dispersion analysis but  the normalisation of the mean scaled profile will be biased high.  This is  further discussed in Sec.~\ref{sec:compdatsim} and Sec.~\ref{sec:scaobs}. 

We now turn to the question of the variation of the pressure profile normalisation with mass. From the definition of $P_{500}$, any deviation from the standard self-similar scaling will appear as a variation of the scaled profiles $p (x)$ with mass, $p(x)\equiv p(x, \Mv)$. It will also translate into a non-standard slope $\alpha_{\rm MY_{\rm X}}$ for the \MY\ relation.   From Eq.~\ref{eq:pm} we expect that $p (x)$ increases slightly with mass as $\YX/\Mv^{5/3}$, i.e as  $ \Mv^{\alpha_{P}}$ with:
\begin{equation}
\alpha_{\rm  P} = \frac{1}{\alpha_{\rm MY_{\rm X}}} - \frac{5}{3}=0.12
\label{eq:alphap0}
\end{equation}
for the best fitting slope $\alpha_{\rm MY_{\rm X}}=0.561$ (Eq.~\ref{eq:mynstd}). We show in the left-top panel of Fig.~\ref{fig:pscmass} the scaled profiles colour coded as a function of $\TX$. There is some indication that hotter (thus more massive) clusters lie above cooler systems. To better quantify the variation with mass, the right-top panel  of the figure shows the variation with $\Mv$ of the scaled pressure, $p(x)$, for different scaled radii, $x=r/\Rv$. At each radius, we fitted the data with a power law $p (x) \propto \Mv^{\alpha(x)}$. At all radii, the slope  $\alpha(x)$ is consistent with the expected $0.12$ value (Fig.~\ref{fig:pscmass}, right top panel), and the mean slope is $0.10\pm 0.02$. The pivot of the power law, where the pressure  equals the average scaled value,  $p(x)= \mathbb{p}(x)$, is about  $\Mv\sim3 \times10^{14}\,\msol$. In a first approximation,  the mass dependence of the scaled profiles can then be modelled by:
\begin{equation}
p (x, \Mv)  =  \mathbb{p}(x)\,\left[\frac{\Mv}{3 \times10^{14}\,{\rm h^{-1}_{70}}\,\msol}\right]^{\alpha_{\rm  P}=0.12 } 
\label{eq:pcors}
\end{equation}
\noindent where $\mathbb{p}(x)$ is the average scaled profile derived in Sec.~\ref{sec:pscm}.

However, there is some indication  that the  mass dependence of the profiles is actually more subtle than a global normalisation variation. The variation in slope of the power law fits shows that the mass dependence  decreases with radius, with $\alpha(x)=0.22\pm0.16$ at $r= 0.1\Rv$ and $\alpha(x)=-0.01\pm0.16$, consistent with zero at $\Rv$. In other words, the departure from standard scaling, likely to be due to the effects of non-gravitational processes, becomes less pronounced as we move towards the cluster outskirts, behaviour that was also noticed in the entropy profiles \citep{nag07,pra09b}.  Note, however, that the mass dependence is weaker for the pressure than for the entropy: the pressure slopes are about two times smaller than those of the entropy \citep[Fig.~\ref{fig:pscmass} and ][their Fig 3]{pra09b}. The variation of $\alpha(x)$ with x can be adequately represented by the analytical expression, $\alpha(x)=\alpha_{\rm P} + \alpha_{\rm P}^{\prime}(x)$, with:
\begin{equation}
\alpha_{\rm P}^{\prime}(x)  =  0.10  - (\alpha_{\rm P} + 0.10) \frac{ (x/0.5)^{3  }}{ 1.+(x/0.5)^{3}}
\label{eq:alphap1}
\end{equation}
\noindent yielding to the more accurate, although more complex, model for the scaled profiles:
\begin{equation}
p (x, \Mv)  =  \mathbb{p}(x)\,\left[\frac{\Mv}{3 \times10^{14}\,{\rm h^{-1}_{70}}\,\msol}\right]^{\alpha_{\rm P} + \alpha_{\rm P}^{\prime}(x) } .
\label{eq:pcor}
\end{equation}

We then compared to the results obtained using $\Mv$ derived from the self-similar \MY\ relation with slope 3/5 (Eq.\ref{eq:mystd}). The scaled profiles are plotted in the bottom panel of Fig.~\ref{fig:pscmass}. In this case, we do not expect  any dependence of $p (x)$ with $\Mv$, and this is indeed the case:  the slopes $\alpha_{P}(x)$ are consistent with zero at all radii  (right bottom panel).   The dispersion in scaled profiles is also smaller (see Fig.~\ref{fig:pscdyn} bottom panel). In that case, the dispersion is only due to structural variations, while  in the non-standard case, the mass dependence of $p(x)$ also contributes to the dispersion.

 %================================================================================
%================================================================================
   \begin{figure}[t]
\centering
\includegraphics[ width=\columnwidth, keepaspectratio]{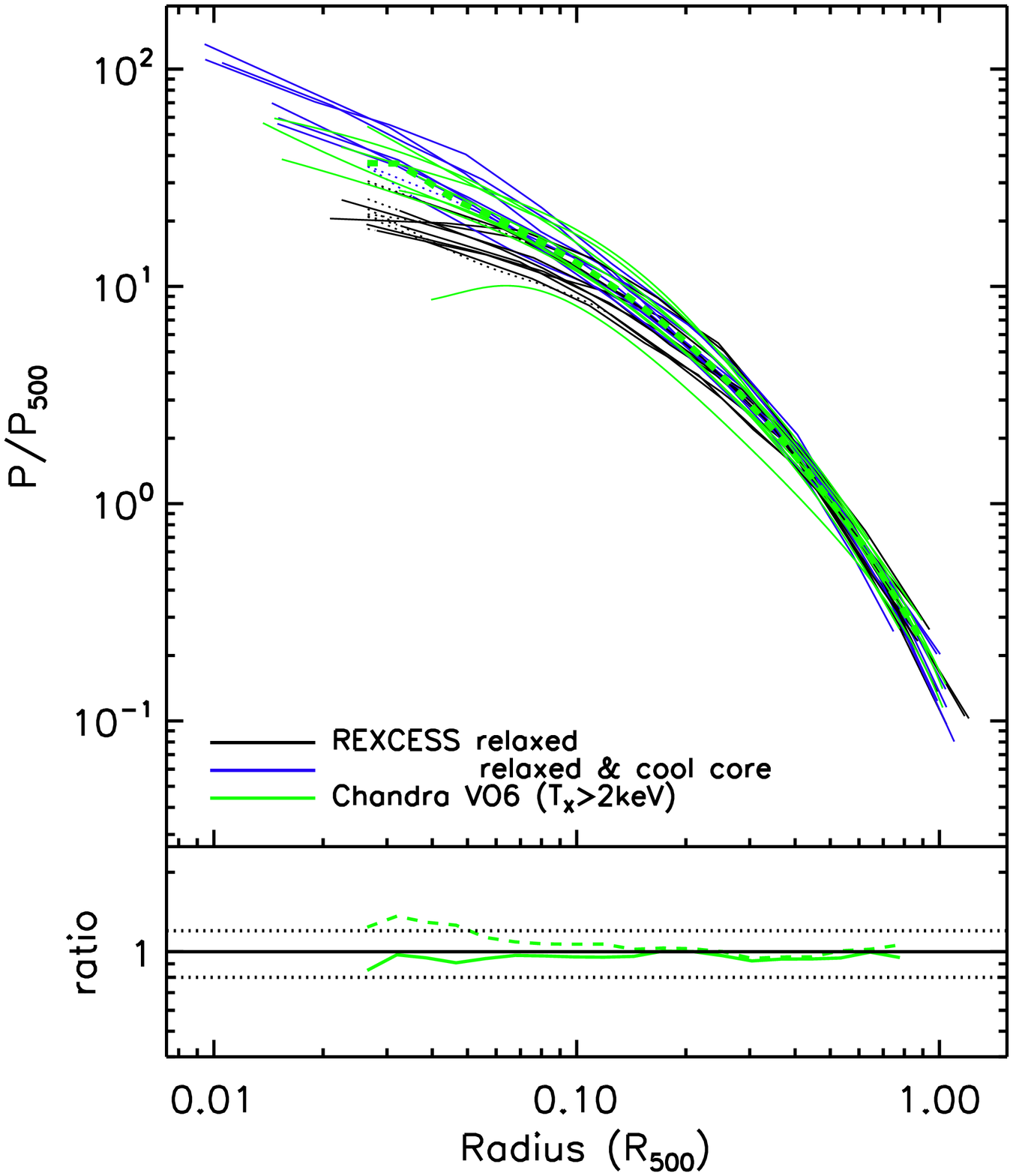} 
 \caption{\footnotesize The scaled pressure profiles (green lines) derived from \citet{vik06a} \chandra\ data on relaxed clusters compared to the scaled profiles of the \rexcess\ sample excluding morphologically disturbed clusters (same colour code as in Fig.~\ref{fig:pscdyn}).   The thick green dotted line is the average \chandra\ profile. Bottom panel: ratio of that average {\it Chandra} profile to that of \rexcess\ for all morphologically undisturbed objects (dotted line) or only cool core clusters (full line).}
 \label{fig:psccha}  
\end{figure}
%================================================================================
%================================================================================

\subsection{Comparison to \chandra\ results for relaxed clusters}
\label{sec:compcha}

In Figure \ref{fig:psccha},  we plot the pressure profiles presented in \citet{nag07}, derived from  \chandra\ data analyzed  by \citet{vik06a}.  We only consider clusters with measured $\Mv$ values, excluding MKW4 ($T=1.4\keV$) and A2390 ($z=0.23$) which fall outside the $\TX$ and $z$ range of \rexcess, respectively.  We used the published $\Mv$ values, derived from the hydrostatic equilibrium (HSE) equation, and  computed the pressure from the best fitting parametric models of the density and temperature profile given in \citet{vik06a}, in the radial range of the observed temperature profile.   Since the \chandra\ data set only contains relaxed clusters,  they are compared to the \rexcess\  profiles excluding  morphologically disturbed objects.  

 All \chandra\ profiles, except one, lie within the range of the \rexcess\ profiles.  The bottom panel of Fig.~ \ref{fig:psccha}  shows the ratio of the average  \chandra\ profile to the average \rexcess\ profile. The agreement between the average profiles, both in shape and normalisation, is nearly  perfect beyond the core, where the dispersion of the scaled profiles is lower.   However, on average, the \chandra\  profiles  are slightly more peaked towards the centre (dotted line in  bottom panel of Fig.~ \ref{fig:psccha}) and have a smaller dispersion than the `relaxed' \rexcess\ clusters.  Better agreement is found with the average \rexcess\ profile for cool core clusters (full line in bottom panel of Fig.~ \ref{fig:psccha}). This is not surprising, since all clusters in the \chandra\ data set  present the central temperature drop characteristic of cool core clusters. 

This good agreement is an indication of the robustness of scaled pressure profile  measurements with current X-ray satellites. The comparison also illustrates the importance  of considering a representative cluster sample to  measure the average profile and dispersion in the core. 

%-----------------------------------------------------------------------------------------------------------------------------------------------------------------------------------
%-----------------------------------------------------------------------------------------------------------------------------------------------------------------------------------
\section{Comparison with numerical simulations}
\label{sec:simul}

\subsection{The data set}

We consider three large samples of simulated clusters at redshift zero extracted from  $\Lambda$CDM  cosmological N-body/hydrodynamical simulations  ($\Omega_{\rm M} = 0.3$, $\Omega_\Lambda= 0.7$). The data set includes the samples from \citet[][hereafter BO]{bor04}, \citet[][PV]{pif08} and \citet[][NA]{nag07}. All simulations include treatment of radiative cooling, star formation, and energy  feedback from supernova  explosions.  The three simulated data sets are fully independent and  derived using  different  numerical schemes and implementations of the gas physics (see references above for full description). This allows us to check the robustness of the theoretical predictions of the pressure profiles by comparing  the three simulated data sets. The  fact that the NA simulation was undertaken on a mesh-based Eulerian code, while the PV and BO simulations were derived from particle-based Lagrangian codes is  particularly relevant, considering some well known cluster-scale discrepancies between the numerical approaches, such as is seen in the  entropy profiles \citep[see, e.g.,  ][and references therein]{voi05, mit09}. The star formation algorithm and the SN feedback model are also quite different both in implementation and in feedback efficiency.

In order to avoid comparison with inappropriately low mass objects we impose the \rexcess\ lower mass limit $\Mv \geq  10^{14} \, M_{\odot}$, leading to a final number of simulated clusters of 93, 88, and 14 for the BO, PV, and NA samples, respectively.  We computed the pressure profile for each cluster using the mass-weighted gas temperature, since the deprojection of the observed profile takes into account the spectroscopic bias (Sec.~\ref{sec:data}). The assumed baryon densities are $\Omega_{\rm b}=0.039, 0.049,0.043$ for the BO, PV, and NA samples, respectively.  The assumed baryon fraction, $f_{b}= \Omega_b / \Omega_m$ has a direct impact on the gas density and thus pressure profile at a given total mass.  We thus corrected the  gas profiles by the ratio between the  assumed $f_{b}$ value and the WMAP5 value \citep{dun09}  for each sample. To scale each individual pressure profile we consider both the  `true' $\Rvs$ and $\Mvs$ values and the hydrostatic values $\Rvh$ and $\Mvh=M^{\rm HSE} (< \Rvh)$. The former are derived from the total mass distribution in the simulation. The latter was derived from the gas density and temperature profiles and the hydrostatic equilibrium equation, using the same procedure for all clusters. As in previous work \citep[e.g.,][and references therein]{pif08}, we find that $\Mvh$ underestimates the true mass. We find a mean bias for the whole sample of $-13$ per cent with a dispersion of $\pm16$ per cent; the average bias estimated for the  different simulations agrees within a few percent at all radii larger than $0.1\Rv$. 

\subsection{Comparison of numerical simulations}
\label{sec:compsimsim}

%================================================================================
%================================================================================
\begin{figure}[tp]
\centering
\includegraphics[ width=\columnwidth, keepaspectratio]{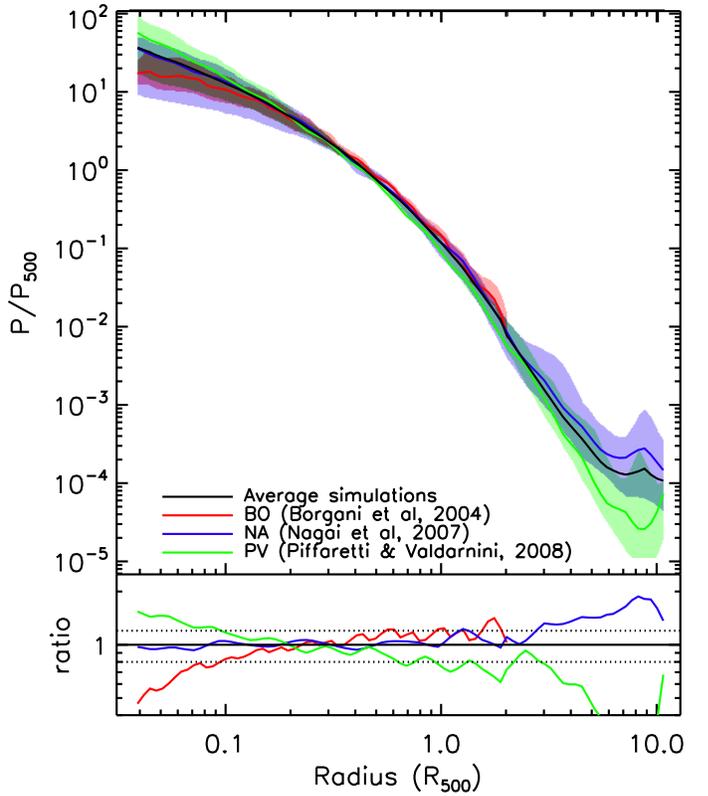}
 \caption{\label{simvssim}  \footnotesize The scaled pressure profiles derived from numerical simulations of \citet{bor04} (pink),  \citet{nag07} (blue) and \citet{pif08} (green). Black line: overall average profile (see text).  coloured lines: average profile for each simulation with the coloured  area corresponding  to the dispersion around it.  Bottom panel: ratio of each simulation average profile to the overall average profile.   } 
\end{figure}
%================================================================================
%================================================================================

We derive the average scaled profile for each simulation, and the dispersion around it,  from the median value and 16 and 84 percentiles of the scaled pressure distribution  at a given scaled radius. We also compute an average simulation profile. Since the average profile computed from the total sample would be biased by the number of objects in the largest data set, we average the three mean profiles from each simulation data set, and calculate the dispersion from all available profiles.
The results derived using the true mass are shown in Fig.~\ref{simvssim}. 

Taking into account that the profiles vary by more than 5 orders of magnitude from the cluster centre to the outskirts, the agreement between the three simulations is exceptionally good. The profiles agree within $20 \%$ between $\sim 0.1$ and $\sim 3 \,\Rvs$ (Fig. \ref{simvssim} lower panel). As expected, larger differences are found in the core, where non-gravitational processes are more important and where the differences in their implementation in the codes will become more evident. 
The BO profiles are available only up to the 'virial' radius, $\simeq 2.03\,\Rvs$ but the PV and NA profiles are traced  up to $10 \,\Rvs$, where they deviate significantly, but  still agree within the dispersion. However,  the differences are sytematic with the PV profiles lying  below the NA profiles. This may hint at a difference in the way in which Lagrangian and Eulerian codes behave in the IGM-WHIM regime.  Note also the flattening of the pressure profile in the outskirts, around  $5 \,\Rvs$, which is likely to define the actual boundary of the cluster, where it meets the intergalactic medium. In the following we will use this boundary  to compute the total integrated SZ signal, $\YSZ$.  In spite of the difference in the pressure in the outskirt,  there is good agreement on $\YSZ$ between the simulations: the SZ signal within $5\,\Rv$ computed from the average PV and NA profiles differ by $-15\%$, and $+9\%$, respectively, from the value computed using the average simulation profile. 

%================================================================================
%================================================================================
\begin{figure}[tp]
\centering
\includegraphics[ width=\columnwidth, keepaspectratio]{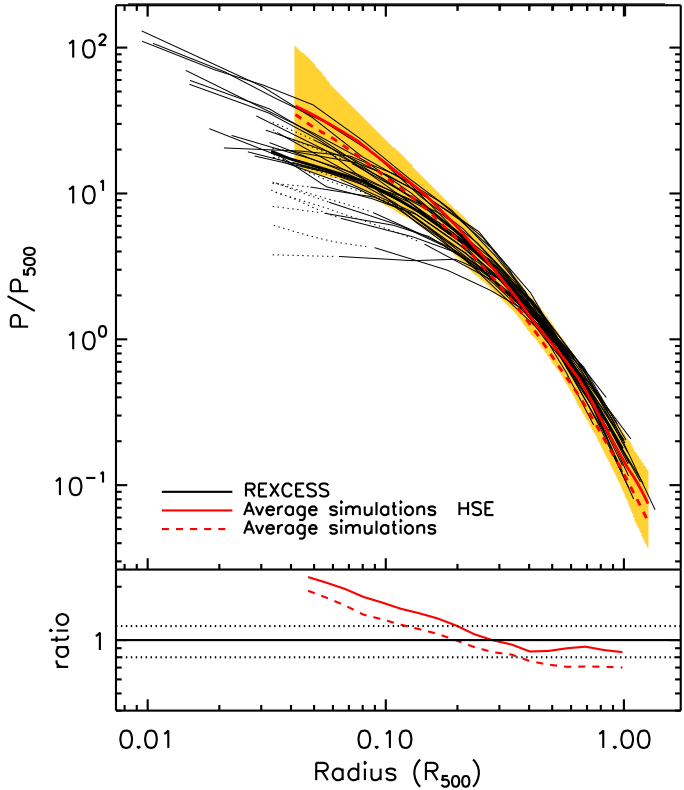}
 \caption{\label{fig:datvssim} \footnotesize  Comparison of the \rexcess\ scaled profiles with the prediction of numerical simulations. Black lines: \rexcess\ data (as in Fig.~\ref{fig:pscdyn}). Thick black line: average \rexcess\ scaled profile. Red line: average simulation profile and dispersion around it (orange area) using the hydrostatic mass. Dotted red line: same using the true mass. Bottom panel: ratio of these average simulation profiles to the \rexcess\ average profile.  } 
\end{figure}
%================================================================================
%================================================================================

\subsection{Comparison of \rexcess\ profiles with simulations}
\label{sec:compdatsim}

Figure~\ref{fig:datvssim} compares the observed scaled profiles with the prediction of  the simulations.
We first consider the simulated profiles scaled using the hydrostatic quantities $\Rvh$ and $\Mvh$, since the observations rely on hydrostatic mass estimates. Note that we used the \MY\ relation calibrated from a sample of relaxed clusters, while for the simulations we used  $\Rvh$ and $\Mvh$ for the whole sample.  However we checked that, when considering only relaxed clusters, the median bias on $\Mv$ changes by only $2\%$, the main effect being a factor of 2 decrease in its dispersion.

The simulation prediction and the \rexcess\ data agree well in the external part ($r\gtrsim 0.2\Rv$), with the observed profiles  lying within the dispersion around the average simulation profile (Fig.~\ref{fig:datvssim}).  
 Remarkably, the observed and simulated average profiles are parallel  above $0.4\Rv$ (i.e they have the same shape), with a normalisation offset of only $\sim 10\%$  (Fig.~\ref{fig:datvssim}, bottom panel). The slight underestimate of the pressure in the simulations is similar to the offset observed for the \MY\ relation and may be due, at least in part,  to over-condensation of hot gas in the cold dense phase \citep[see discussion in][]{app07}.  As we move towards the centre, the agreement progressively  degrades, the simulations  predicting  more peaked profiles than those observed (Fig.~\ref{fig:datvssim} bottom panel).  This behaviour was also noticed by \citet{nag07}  when comparing their simulations with \chandra\ relaxed clusters, and it is also observed for the temperature profiles \citep[see][]{pra07}.  As mentioned above, the core properties are most sensitive to non-gravitational processes and these discrepancies  are again likely to reflect  the fact that modelling of the processes is still inadequate. 

The average simulation profile derived using the true mass for each simulated cluster is also shown in the figure (dotted lines). As compared to the scaling based on $\Rvh$ and $\Mvh$, the scaled profile of each cluster is translated  to the left and to the bottom in the log-log plane. The average profile  lies below the profile based on the hydrostatic values,  as expected from the mean bias between $\Mvs$ and $\Mvh$.  The offset with the observed profile in the outer region becomes more significant, about $30\%$. 

In conclusion, there is an excellent agreement in shape between the simulated and observed profiles for the cluster outer regions, which is the most relevant aspect for the $\YSZ$ estimate. The better agreement in normalisation with the simulations when using the hydrostatic mass suggests that the hydrostatic X-ray masses used to scale the observed profiles are indeed underestimated.  

%-----------------------------------------------------------------------------------------------------------------------------------------------------------------------------------
%-----------------------------------------------------------------------------------------------------------------------------------------------------------------------------------
\section{The universal pressure profile}
\label{sec:puniv}

As pointed out by \citet{nag07}, an analytic cluster pressure profile model is useful both for analysis of SZ observations and for theoretical studies. Of prime interest is a model for the average scaled profile of the entire cluster population. For nearby clusters it can be derived from the present data, the \rexcess\ sample being a representative sample. 

We considered the generalized NFW (GNFW) model proposed  by \citet{nag07}:
\begin{equation}
 \mathbb{p}(x) = \frac{P_{0} } { (c_{500}x)^{\gamma}\left[1+(c_{500}x)^\alpha\right]^{(\beta-\gamma)/\alpha} } 
\label{eq:pgnfw}
\end{equation}
\noindent The parameters $(\gamma,\alpha,\beta)$ are respectively the central slope ($r\ll \rs$), intermediate slope ($r \sim \rs$) and outer slope  ($r\gg \rs$), where $\rs= \Rv/c_{500}$, and they are highly correlated with $\rs$. In order to constrain the parameters, it is essential to consider a wide radial range, including both the core ($r<0.1\Rv$) and the cluster periphery ($r>\Rv$). In particular, $\beta$ remains essentially unconstrained when considering only data within  $r<\Rv$, resulting in large uncertainties in the profile model beyond $\Rv$ and thus on the corresponding integrated SZ signal.  

\begin{figure}[tp]
\centering
\includegraphics[ width=\columnwidth, keepaspectratio]{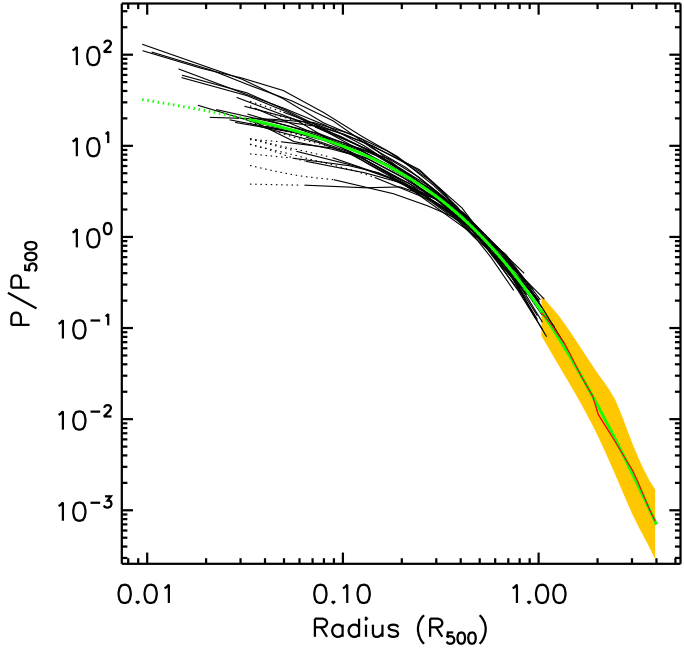} 
 \caption{\label{fig:puniv} \footnotesize GNFW model of the universal pressure profile (green line). It is derived by fitting the observed average scaled profile  in the radial range $[0.03$--$1]\Rv$, combined with the average simulation profile  beyond $\Rv$  (red line).  Black lines:  \rexcess\ profiles.  Orange area: dispersion around the average simulation profile.} 
\end{figure}

Taking  advantage of the good agreement between observations and simulations in the outer cluster regions, we thus defined an hybrid average profile, combining the profiles from observations and simulations.  It is defined by the  observed average scaled profile in the radial range $[0.03$--$1]\Rv$ derived in Sec.~\ref{sec:pscm}  and  the average simulation profile in the $[1$--$4]\Rv$ region. For the simulations, we used the profile based on the hydrostatic quantities and renormalised it by $+10\%$ to correct for the observed offset with the observations at $r>0.4\Rv$. 
We fitted this hybrid profile with the GNFW model in the log-log plane, weighting the `data' points according to the dispersion. The best fitting model is plotted in Fig.~\ref{fig:puniv}, with parameters: 

{\small
\begin{equation}
[P_{0} ,c_{500},\gamma,\alpha,\beta] =  [8.403\,{\rm h_{70}^{-3/2}},1.177,0.3081,1.0510,5.4905]
\label{eq:pargnfw}
\end{equation}
}

\noindent Using the dimensionless  `universal'  profile, $\mathbb{p}(x)$ (Eq.~\ref{eq:pgnfw} and Eq.~\ref{eq:pargnfw}), and taking into account the mass dependence established in Sec.~\ref{sec:pscmass}, we can describe the physical pressure profile of clusters as a function of mass and redshift (assuming standard evolution):
\begin{eqnarray}
\label{eq:puniv}
P(r) & =  & P_{500} \left[\frac{\Mv}{3 \times 10^{14}\,{\rm h^{-1}_{70}}\,\msol}\right]^{ \alpha_{\rm P}+\alpha^{\prime}_{\rm P}(x)}~ \mathbb{p}\left(x\right) \\
 & = &1.65\times10^{-3}\,h(z)^{8/3}\,\left[\frac{\Mv}{3 \times 10^{14}\,{\rm h^{-1}_{70}}\,\msol}\right]^{2/3 +\alpha_{\rm P}+\alpha^{\prime}_{\rm P}(x)} \nonumber \\
& & {} \times \mathbb{p}\left(x\right)~~{\rm h_{70}^{2}\,\keV\,cm^{-3}} \nonumber
\end{eqnarray}
\noindent with $x=r/\Rv$, $\alpha_{\rm P}$ and $\alpha^{\prime}_{\rm P}(x)$ from Eq.~\ref{eq:alphap0} and Eq.~\ref{eq:alphap1}, and $ \mathbb{p}(x)$ from Eq.~\ref{eq:pgnfw} with parameters from Eq.~\ref{eq:pargnfw}.
The second term in the mass exponent, $\alpha_{\rm P}$, corresponds to a modification of the standard self-similarity (i.e., the steeper mass dependence of the profile), while  the third term, $\alpha^{\prime}_{\rm P}(x)$ (Eq.~\ref{eq:alphap1}), introduces a break in self-similarity (i.e., a mass dependence of the shape).  The latter is a second order effect, which  can be neglected in first aproximation. 

%================================================================================
%================================================================================
%================================================================================
%================================================================================

We also fitted each individual observed  cluster profile with the GNFW model, fixing the $\beta$ value to that derived above (Eq.~\ref{eq:pargnfw}). The best fitting parameters are listed in Appendix~\ref{ap:pnfw}, where we also provide plots of each individual cluster profile with its best fitting model.

%-----------------------------------------------------------------------------------------------------------------------------------------------------------------------------------
%-----------------------------------------------------------------------------------------------------------------------------------------------------------------------------------
\section{Integrated Compton parameter scaling relations}
\label{sec:ysz}
\subsection{Definitions and method} 

In this section we discuss scaling relations directly relevant for SZE studies. We will consider the volume integrated Compton parameter $Y$, for both cylindrical and spherical volumes of integration. The spherically integrated quantity, $Y_{\rm sph}(R)$, proportional to the gas thermal energy, is defined as:
 \begin{equation}
 Y_{\rm sph}(R)=\frac{\sigma_{\rm T}}{m_{\rm e} c^2}\,\int_{0}^{R} 4\pi P(r)r^2dr
 \label{eq:ysph}
 \end{equation}
\noindent and the  cylindrically integrated quantity, $Y_{\rm cyl}(R) = \YSZ D_{A}^2$, directly related to the SZ signal within an aperture $\theta=R/D_{A}$, is:
 \begin{eqnarray}
 \label{eq:ycyl}
Y_{\rm  cyl}(R)&=&\frac{\sigma_{\rm T}}{m_{\rm e} c^2}\,\int_{0}^{R} 2\pi r dr \int_{r}^{R_{\rm b}} \frac{2\,P(r')r'dr'}{\sqrt{r'^2 -r^2}}  \\
& = &  Y_{\rm sph}(R_{\rm b})-\frac{\sigma_{\rm T}}{m_{\rm e} c^2}\,\int_{R}^{R_{\rm b}} 4\pi\,P(r)\sqrt{r^2 -R^2}rdr  \nonumber
 \end{eqnarray}
\noindent where $R_{\rm b}$ is the cluster radial extent. In the following, we adopt $R_{\rm b}=5\Rv$, as suggested by numerical simulations (Sec.~\ref{sec:compsimsim}). Note that the total SZ signal is then equivalently $Y_{\rm  sph}(5\Rv)$ or $Y_{\rm  cyl}(5\Rv)$.

For each cluster, the  spherically integrated Compton parameter can be readily computed from the observed pressure profile. The $Y_{\rm sph}$ scaling relations can then be directly derived from the data for integration radii up to $\Rv$,  the observed radial range. They are presented below in Sec.~\ref{sec:scaobs}. Such a derivation is not possible for $Y_{\rm cyl}$ (or the total $Y_{\rm SZ}$ signal):  it involves  integration  along the line of sight up to $R_{\rm b}=5\Rv$, i.e., beyond the observed radial range. However, using the universal pressure profile, we can compute the volume integrated Compton parameter, $Y$, for any region of interest, and derive the corresponding scaling relations (presented below in Sec.~\ref{sec:scauniv}). The two approaches give fully consistent results, as shown below. 
 
Finally, for convenience, we also define a characteristic Compton parameter,  $Y_{500}$, corresponding to the characteristic pressure $P_{500}$ (see Appendix~\ref{ap:ss}):
\begin{eqnarray}
 \label{eq:y500}
Y_{500} &= &  \frac{\sigma_{\rm T}}{m_{\rm e}\,c^2}\,\frac{4\pi}{3}\,\Rv^3\,P_{500}   \\
 &= &2.925   \times 10^{-5} h(z)^{2/3} \left[\frac{\Mv}{3 \times 10^{14}\,{\rm h^{-1}_{70}}\,\msol}\right]^{5/3}~~{\rm h_{70}^{-1}\,Mpc^{2}} \nonumber
\end{eqnarray}
\subsection{Observed $Y_{\rm sph}$--$\YX$ and  $Y_{\rm sph}$--$\Mv$ relations} 
\label{sec:scaobs}

The values for $Y_{\rm sph}(R_{2500})$ and $Y_{\rm sph}(R_{500})$, derived from the observed pressure profiles, are given in Table~\ref{tab:pnfw}. $R_{2500}$ is defined as $R_{2500}=0.44\Rv$ from the scaling relations presented in \citet{app05}.  The integration was performed using the MC deconvolved density and model temperature profiles, allowing us to propagate the statistical errors, including that on $\Rv$. We checked that using instead the best fitting GNFW model for each profile gives consistent results within the statistical errors.  Note that the errors on $\Mv$ take into account the statistical errors on the relevant X-ray data, but  not the uncertainties on the \MY\ relation itself. The latter are therefore not included in the  statistical errors on the slope and normalisation of the relations.

 %================================================================================
%================================================================================
\begin{figure}[tp]
\centering
\includegraphics[ width=\columnwidth, keepaspectratio]{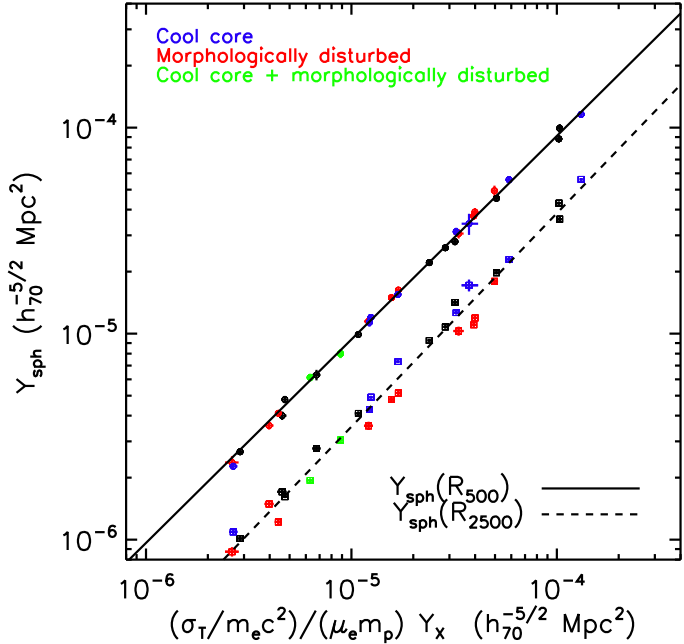} 
 \caption{\label{fig:yszyx} \footnotesize The $Y_{\rm sph}$--$\YX$ relations from \rexcess\ data.  $Y_{\rm sph}(R)$ is the spherically integrated Compton parameter, within $R_{2500}$ (squares) and $\Rv$ (circles).  $\YX=\Mgv\TX$ is the product of the gas mass within $\Rv$ and the spectroscopic temperature $\TX$. Data points are colour-coded according to cluster dynamical state. Lines: best fitting power law. } 
\end{figure}
%================================================================================
%================================================================================

 Figure~\ref{fig:yszyx} shows the $Y_{\rm sph}$--$\YX$  relations with  $\YX=\Mgv\TX$, together with the best fitting power law. We normalised  $\YX$ by :
\begin{equation}
C_{\rm XSZ} =\frac{\sigma_{\rm T}}{m_{\rm e} c^2} \frac{1}{\mu_{\rm e} m_{\rm p}} =   1.416 \times 10^{-19}~~\frac{{\rm Mpc}^2}{\rm \msol\, \keV}
\label{eq:cyszyx}
\end{equation}
\noindent for $\mu_{\rm e}=1.148$, the mean molecular weight of electrons for a 0.3 solar abundance plasma.   Note that the $Y_{\rm sph}$--$\YX$ relation depends only weakly on the assumed \MY\ relation, via the estimate of $\Rv$ only. For some clusters, the computation of $Y_{\rm sph}(R_{500})$ requires extrapolation: by more than $20\%$ for 8 clusters and, in the worst case, RXC J2157.4-0747, the profile of which is measured only up to $R_{\rm det}\sim0.6 \Rv$, $Y_{\rm sph}(\Rv)$ is larger by a factor $1.8$ than the value within $R_{\rm det}$. However, the  best fitting $Y_{\rm sph}(R_{500})$--$\YX$ relation is stable to the inclusion or exclusion of clusters requiring extrapolation, the best fitting parameters being consistent within the errors. 

As mentioned in the introduction, the $Y_{\rm sph}$--$\YX$  relation  depends on the internal cluster  structure (Eq.~\ref{eq:yszyx}). For $Y_{\rm sph}(R_{2500})$,  we obtained:
\begin{equation}
Y_{\rm sph}(R_{2500}) = 10^{-0.272\pm0.097} \left[\frac{C_{\rm XSZ}\YX}{ {\rm h_{70}^{-5/2}\,Mpc^2}}\right]^{1.036\pm0.020}\!\!\!\!{\rm h_{70}^{-5/2}\,Mpc^2}
\end{equation}
The best fitting slope is slightly greater than one (a $2\sigma$ effect), reflecting the stronger mass dependence of the pressure profile in the centre ($r<R_{2500}$) as compared the expectation from the \MY\ relation (Fig.~\ref{fig:pscmass} and Sec.~\ref{sec:pscmass}). The intrinsic dispersion is $\sigma_{\rm log10, Y}= 0.054 \pm0.006$, with the morphologically disturbed clusters lying  below the mean relation and the relaxed clusters lying above it,  a consequence of the  shallower profile of the former as compared to the latter (Fig.~\ref{fig:pscdyn}).   When we move to $Y_{\rm sph}(\Rv)$, the best fitting slope ($1.003\pm0.008$) becomes consistent with unity, i.e  the shape variation with mass, when averaged within $\Rv$,  has essentially no effect (see also below).  The intrinsic dispersion is  no longer measurable, the dispersion is consistent with that expected from the statistical errors. This is a direct consequence of the high similarity of the pressure profiles beyond the core ($r \gtrsim0.2 \Rv$), while the core  typically contributes by less than $10\%$ to  $Y_{\rm sph}(500)$ (see below and Fig.~\ref{fig:punivcomp}). 
Fixing the slope to one, the best fitting normalisation gives:
\begin{equation}
\frac{Y_{\rm sph}(\Rv) }{C_{\rm XSZ}\,\YX} =   0.924\pm0.004 
\label{eq:yszyx500}
\end{equation}
Note that this ratio is nothing more than the ratio, $T_{\rm mg}/\TX$, of the gas mass weighted temperature to $\TX$. It is less than unity, as  found in other studies \citep{vik06a}, and as expected for decreasing temperature profiles. 

%================================================================================
%================================================================================
\begin{figure}[tp]
\centering
\includegraphics[width=\columnwidth, keepaspectratio]{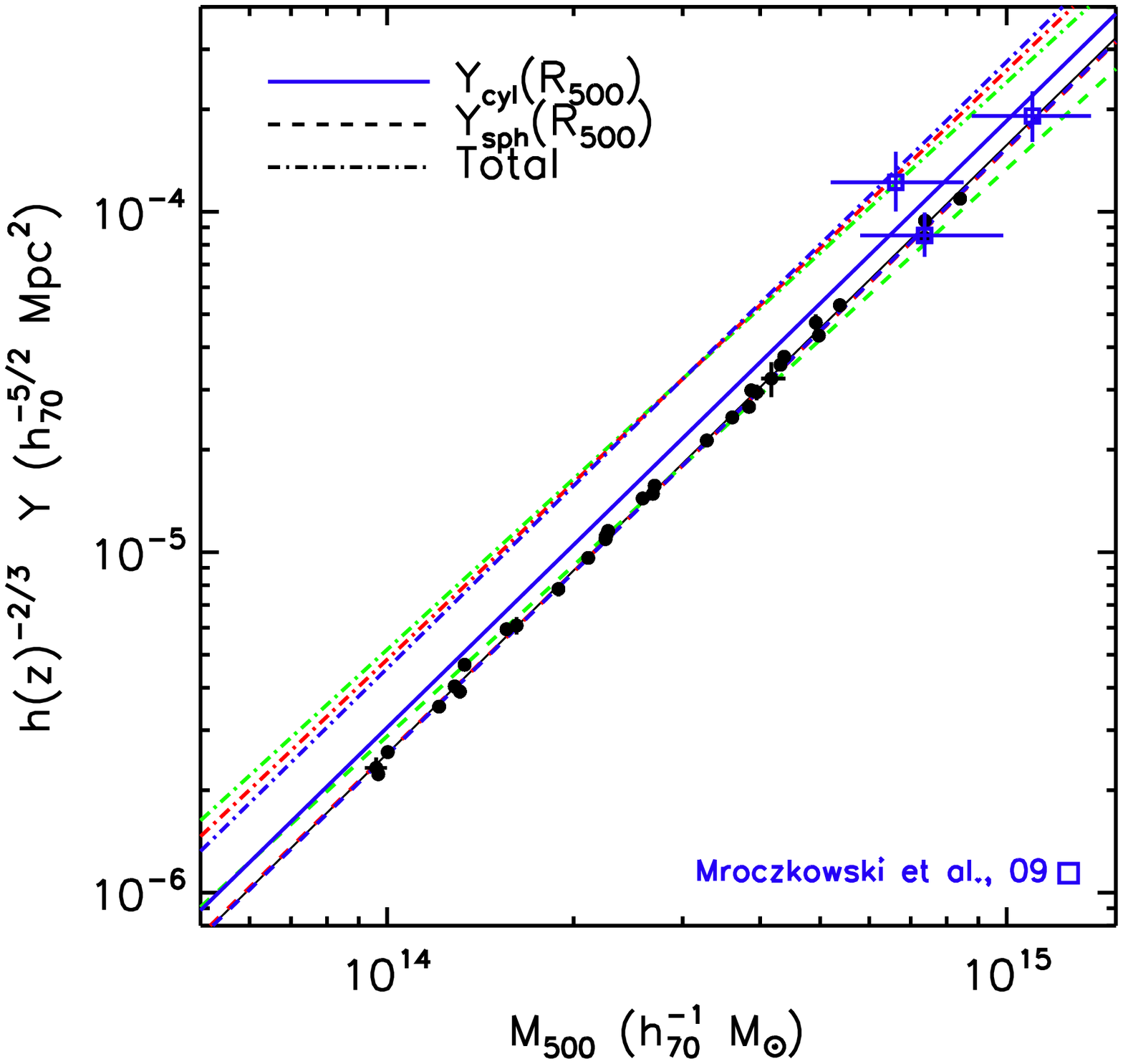} 
 \caption{\footnotesize The $Y_{\rm SZ}$--$\Mv$ relations.  Thin black line:  power law  $Y_{\rm sph}(\Rv)$--$\Mv$ relation best fitting the \rexcess\ data  (black points). Lines:  $Y_{\rm sph}(\Rv)$--$\Mv$ relation (dashed line),  $Y_{\rm cyl}(\Rv)$--$\Mv$ relation (relation between  $Y_{\rm SZ}D_{\rm A}^2$ within a  $\Rv$ aperture and $\Mv$; full line) and  $Y_{\rm cyl}(5\Rv)$--$\Mv$ relation (relation between the total $Y_{\rm SZ}D_{\rm A}^2$ signal and $\Mv$; dash-dotted line) derived from the universal GNFW  scaled pressure profile and for different $\Mv$ scaling: standard self-similar scaling ($\alpha=5/3$; green), modified scaling taking into account the non-standard slope of the \MY\ relation ($\alpha=1.78$; blue), and further taking into account the break of self-similarity of the pressure profile shape (see text, red). Blue squares:  ($Y_{\rm cyl}(\Rv),\Mv$) measurements for  3 clusters: from top to bottom, A1835 $(z=0.25)$, A1914 $(z=0.17)$,  and CL J1226.9+3332 $(z=0.89)$. They were derived by \citet{mro09} from a joint analysis of SZA and X-ray observations using a GNFW model.} 
 \label{fig:ym} 
\end{figure}
%================================================================================
%================================================================================

Figure~\ref{fig:ym} shows the $Y_{\rm sph}(\Rv)$--$\Mv$ data together with the best fitting relation:

{\small
\begin{equation}
h(z)^{-2/3} Y_{\rm sph}(\Rv)= 10^{-4.739\pm0.003} \left[\frac{\Mv}{3 \times 10^{14}\,{\rm h^{-1}_{70}}\,\msol}\right]^{1.790\pm0.015}\!\!\!\!\!\!\!{{\rm h_{70}^{-5/2}\,Mpc}^2}
\label{eq:mysz500}
\end{equation}
}

\noindent Since $\Mv$ is derived from the \MY\ relation, this expression does not contain more information than the  $Y_{\rm sph}(\Rv)$--$\YX$ relation, combined with the calibration of the  \MY\ relation. As expected, the normalisation and slope are consistent with that obtained by combining Eq.~\ref{eq:mynstd} and Eq~\ref{eq:yszyx500},  and, similar to the $Y_{\rm sph}(\Rv)$--$\YX$ relation, the scatter is consistent with the statistical scatter. 

With such a study, based on a mass proxy,  we cannot assess the intrinsic scatter of the `true' $Y_{\rm sph}(\Rv)$--$\Mv$ relation. However, we emphasize that our study does show, from the low scatter of the $Y_{\rm sph}(\Rv)$--$\YX$,  that variations in pressure profile shapes do not introduce an extra scatter into the  $Y_{\rm sph}(\Rv)$--$\Mv$ relation as compared to that of the $\YX$--$\Mv$ relation. Since  the latter was derived from  hydrostatic mass estimates  using relaxed objects, the above $Y_{\rm sph}(\Rv)$--$\Mv$ is expected to differ from the `true' $Y_{\rm sph}(\Rv)$--$\Mv$ by the offset between the `true' mass and the  hydrostatic mass for {\it relaxed} objects.  

\subsection{Scaling relations from the universal pressure profile}
\label{sec:scauniv}

\subsubsection{$Y_{\rm sph}$--$\Mv$ and  $Y_{\rm SZ}D_{\rm A}^2$--$\Mv$ relations}

Let us first consider $Y_{\rm sph}$ derived from the universal pressure profile. Combining Eq.~\ref{eq:ysph}, \ref{eq:puniv} and \ref{eq:y500}:
%{\small
\begin{eqnarray}
\label{eq:mysph0}
Y_{\rm sph}(R)  & = & Y_{500}  \left[\frac{\Mv}{3 \times10^{14}\,{\rm h^{-1}_{70}}\,\msol}\right]^{ \alpha_{\rm P}} \\
& & {} \times \int_{0}^{x}3f(u,\Mv)\,\mathbb{p}(u)\,u^2\, du \nonumber
\end{eqnarray}
%}
with $f(u,\Mv)=(\Mv/3 \times10^{14}\,{\rm h^{-1}_{70}}\,\msol)^{\alpha^{\prime}_{\rm P}(u)}$.  This term in the integral reflects the break of self-similarity in the pressure profile (Sec.~\ref{sec:puniv}).  Neglecting this effect, the corresponding $Y_{\rm sph}$--$\Mv$ relation, for any integration radius, is a power law of slope $\alpha=5/3+ \alpha_{\rm P}=1/\alpha_{\rm MY_{\rm X}}$ (Eq.~\ref{eq:alphap0} and \ref{eq:y500}). Taking into account this effect, the relation is no  longer a simple power law. Following the behavior of the pressure profiles -- $\alpha^{\prime}_{\rm P}(u)$ decreases with radius or equivalently  the  departure from standard mass scaling becomes less and less pronounced as we move towards the cluster outskirts -- the relation  is expected to become shallower with increasing integration radius, closer to the standard self-similar relation ($\alpha=5/3$).  The relations for various mass scalings can be compared in Fig.~\ref{fig:ym}, for $Y_{\rm sph}(\Rv)$ and $Y_{\rm sph}(5\Rv)$ (i.e the total $Y_{\rm SZ}$ signal). The effect of the self-similarity break is small. In the mass range $\Mv=[10^{14}\msol,10^{15}\msol]$, $Y_{\rm sph}$ varies, as compared to the value computed neglecting this effect,  by $[-7\%,+8\%]$,$[-1\%,+0.5\%]$ and $[+6\%,-6\%]$,  for an integration radius of $R_{2500}$, $\Rv$ and $5\Rv$, respectively. When taking into account the self-similarity break,  the corresponding effective slopes of the $Y_{\rm sph}$--$\Mv$ for that mass range are $1.84$, $1.78$ and $1.73$, as compared to  $1/\alpha_{\rm MY_{\rm X}}=1.78$ ignoring the effect.  The effect is fully negligible for the  $Y_{\rm sph}(\Rv)$--$\Mv$ relation, as found above directly from the data; it is at most equal to the statistical uncertainty  on $1/\alpha_{\rm MY_{\rm X}}=1.78\pm0.06$ (Eq.~\ref{eq:mynstd}) and we will neglect it in the following. 

In that case, and combining Eq.~\ref{eq:mysph0}, \ref{eq:alphap0} and \ref{eq:y500}, the $Y_{\rm sph}$--$\Mv$ relation for an integration radius of  $x\,\Rv$ can be written as:
\begin{equation}
h(z)^{-2/3} Y_{\rm sph}(x\,\Rv)   =   A_{\rm x} \left[\frac{\Mv}{3 \times 10^{14}\,{\rm h^{-1}_{70}}\,\msol}\right]^{\alpha}
\label{eq:mysph}
\end{equation}
where
\begin{eqnarray}
\alpha &=& 1.78;~~~A_{\rm x}\!=\!2.925\times10^{-5}\,I(x)~~{\rm h^{-1}_{70}\,Mpc^2}  \\
I(x)\!&=&\!\int_{0}^{x}3\mathbb{p}(u)\,u^2\,du   
\label{eq:Ix} 
\end{eqnarray}
with $\mathbb{p}(u)$ from Eq.~\ref{eq:pgnfw} and~\ref{eq:pargnfw}. Numerical values for I(x) of particular interest  are $I(1)\!=\!0.6145$ and $I(5)\!=\!1.1037$. The former gives the normalisation of the $Y_{\rm sph}(\Rv)$--$\Mv$ relation, $\log(A_{\rm x})\!=\!-4.745$. It is in excellent agreement ($1\%$ difference)  with the normalisation derived from a direct fit to the data (Eq.~\ref{eq:mysz500}). 
 The latter gives the normalisation, $\log(A_{\rm x})=-4.491$, of the relation for the total $Y_{\rm SZ}D_{\rm A}^2$ signal, assuming a cluster radial extent of $5\Rv$. 

Similarly, the  relation  for the SZ signal within an aperture of $x\,\Rv$ is obtained from Eq.~\ref{eq:ycyl}, \ref{eq:puniv} and \ref{eq:y500}:
\begin{equation}
h(z)^{-2/3} \YSZ(x\,\Rv)\,D_{\rm A}^2   =  B_{\rm x} \left[\frac{\Mv}{3 \times 10^{14}\,{\rm h^{-1}_{70}}\,\msol}\right]^{\alpha}
\label{eq:mysz}
\end{equation}
with
\begin{eqnarray}
\alpha &=& 1.78;~~~B_{\rm x}\!=\!2.925\times10^{-5}\,J(x)~{\rm h^{-1}_{70}\,Mpc^2}  \\
J(x)\!&=&\!I(5)- \int_{x}^{5} 3\, \mathbb{p}(u)\,\sqrt{u^2-x^2}\,u\,du  
\label{eq:Jx} 
\end{eqnarray}
for a cluster extent of $5\Rv$. For an aperture of $\Rv$, $J(1)=0.7398$ or  $\log(B_{\rm x})=-4.665$. The corresponding  $Y_{\rm SZ}D_{\rm A}^2$--$\Mv$ relation is plotted in Fig.~\ref{fig:ym}. We also show measurements for A1835 $(z=0.25)$, A1914 $(z=0.17)$,  and CL J1226.9+3332 $(z=0.89)$, derived by \citet{mro09} from a joint analysis of SZA and X-ray observations using a GNFW pressure profile model.  Although the measurement errors are still large, the consistency with the present scaling relation is an encouraging sign of the validity of our determination of the scaling relations. Since the clusters cover a wide redshift range, it further suggests a standard self-similar evolution, as assumed in Eq.~\ref{eq:mysz}.

Uncertainties on the above relations, that are established combining observational and theoretical data, cannot be assessed rigourously. Rough  estimates of the statistical errors can be derived by combining the errors on the $Y_{\rm sph}(\Rv)$--$\YX$ and \MY\ relations, with the latter largely dominant. This gives $\alpha=1.78\pm0.06$ or $\alpha=1.78\pm0.08$, further adding quadratically the systematic effect of the pressure self-similarity break discussed above.  The logarithmic error on the normalisation at the pivot is $\pm0.024$ ($\pm6\%$).

\subsubsection{Behavior of $Y_{\rm sph}(R)$ and comparison with the isothermal $\beta$--model}

\begin{figure}[tp]
\centering
\includegraphics[ width=\columnwidth, keepaspectratio]{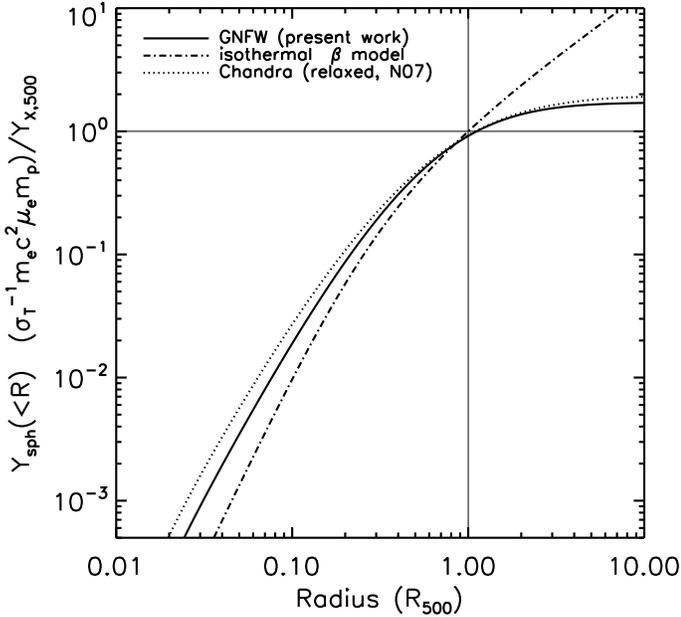} 
 \caption{\label{fig:punivcomp} \footnotesize Spherically integrated Compton parameter, $Y_{\rm sph}(R)$, as a function of scaled integration radius,  normalised to $C_{\rm XSZ}\YX $.  Full line: universal GNFW  scaled pressure profile  (Eq.~\ref{eq:pgnfw} and~\ref{eq:pargnfw}). Dotted line: GNFW model obtained  by \citet{nag07} for \chandra\ relaxed ($T\!>\!5\,\keV$) clusters. Dashed-dotted line:  isothermal $\beta$--model with $\beta=2/3$ and a core radius of $0.2\Rv$.} 
\end{figure}

It is instructive to study in more detail the radial dependence of $Y_{\rm sph}$. $Y_{\rm sph}(R)$ varies with radius as $I(x)$ (Eq.~\ref{eq:Ix} with $\mathbb{p}(u)$ from Eq.~\ref{eq:pgnfw} and \ref{eq:pargnfw}). By construction its normalisation scales with mass as $\YX$. Figure~\ref{fig:punivcomp}  shows the variation of $Y_{\rm sph}(R)$ with scaled integration radius, normalised to $C_{\rm XSZ}\YX$, so that we are effectively probing $Y_{\rm sph}(R)$ at fixed mass.
%================================================================================
%================================================================================
%================================================================================
%================================================================================

At large radii, the integrand in $I(x)$ varies as $\mathbb{p}(u)u^2du \propto u^{-2}$  for an outer slope $\beta\sim5$. As a result,  $Y_{\rm sph}(R)$  converges rapidly beyond $\Rv$ and  the total SZ signal is not very sensitive to the assumption on cluster extent. Assuming a cluster extent of $4\Rv$, $6\Rv$ or even $100\Rv$, rather than $5\Rv$, changes the total signal by only $-2\%$,$+1.3\%$ and $+4\%$,  respectively. On the other hand, the figure shows the dominant contribution of the external regions to $Y_{\rm sph}$: $50\%$ of the contribution to $Y_{\rm sph}(\Rv)$ comes from $R\gtrsim0.53\Rv$ while the region within $0.1\Rv$ and $0.2\Rv$ contributes by only $2\%$ and $9.5\%$ respectively.  This will be even more pronounced for the $Y_{\rm SZ}$ signal (integration within a cylindrical volume).

 We also plot $Y_{\rm sph}(R)$ for the GNFW model obtained  by \citet{nag07} from \chandra\ data (for the corrected parameters, $[ 12.2,1.3,0.4, 0,9, 5.0]$, published by \citealt{mro09}).  It is slightly larger in the centre, as expected from the more peaked nature of the scaled \chandra\ profiles (Sec.~\ref{sec:compcha}). The agreement\footnote{ Note, however, that  \citep{nag07} assumed a standard self-similar mass scaling of the presure profile. The $Y$--$\Mv$ relations derived from their profiles would differ from ours in terms of slope.}  is very good in the outskirts, as it is for the profiles (Fig.~\ref{fig:psccha}), with a slightly higher assymptotic  value due the slightly smaller value of  $\beta$.

We also compare with the result obtained with an isothermal $\beta$--model, with $\beta=2/3$ and a core radius of  $0.2\Rv$ \citep{arn02}. The difference is only $10\%$  at $\Rv$ but the model diverges at high radii. This clearly shows that the total $\YSZ$ signal derived assuming  an isothermal  $\beta$--model  is very sensitive to the assumed extent of the cluster. It will also be always overestimated by such a model, as emphasized by \citet{hal07}. As an illustration, assuming a cluster extent of $2.03\,\Rv$, the top--hat virial radius often used in the litterature, the $\beta$--model gives a total  $Y_{\rm SZ}$ signal $1.7$ higher than the universal pressure profile. 

\subsection{The $Y$--$\LX$ relations}
\label{sec:ylx}
The scaling between the SZ signal and the X--ray luminosity, $\LX$ is an important relation for comparing X--ray surveys such as the \rosat\ All Sky Survey and future or on going SZE surveys, such as the \planck\ survey. The luminosity within $\Rv$ and in the soft-band $[0.1$--$2.4]$~\keV, most relevant for X--ray Surveys, has been estimated for \rexcess\ clusters by \citet{pra09}; here we used the values both corrected and uncorrected for Malmquist bias.   

%================================================================================
%================================================================================
\begin{figure}[t]
\centering
\includegraphics[width=\columnwidth, keepaspectratio]{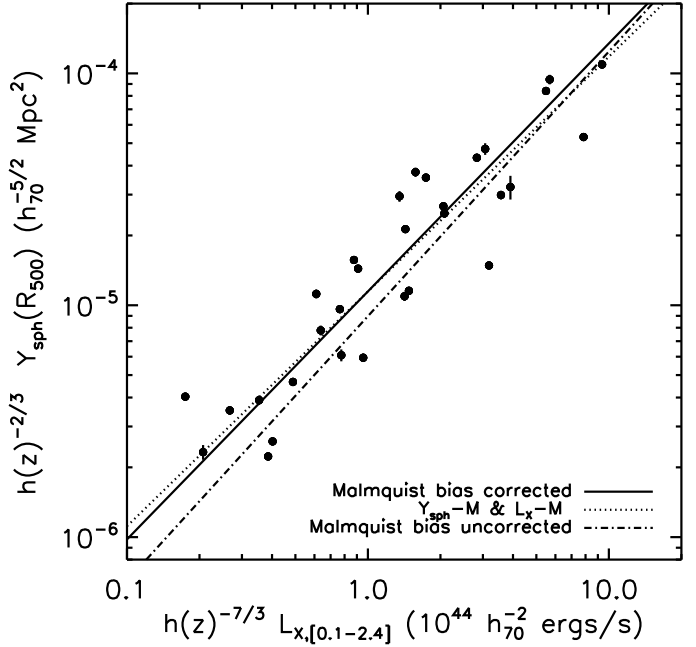} 
 \caption{\label{fig:YLX} \footnotesize The $Y_{\rm sph}(\Rv)$--$\LX$ relations.  $\LX$ is the $[0.1$--$2.4]\,\keV$ luminosity  within $\Rv$. Full line:  power law  relation best fitting the \rexcess\ data for $\LX$ corrected for Malmquist bias  (black points).  Dotted line: relation computed  by combining the $Y_{\rm sph}(\Rv)$--$\Mv$ relation derived from the universal pressure  profile (Eq.~\ref{eq:mysph})  and the  $\LX$--$\Mv$ relation. Dash-dotted Line:  best fitting $Y_{\rm sph}(\Rv)$--$\LX$ for uncorrected  $\LX$.} 
\end{figure}
%================================================================================
%================================================================================

Figure  \ref{fig:YLX} shows the corresponding $Y_{\rm sph}(\Rv)$--$\LX$ relations. We fitted the \rexcess\ data with a power law:
\begin{equation}
h(z)^{-2/3} Y_{\rm sph}(\Rv)   =   C \left[\frac{\LX}{10^{44}\,{\rm h^{-2}_{70}}\,{\rm ergs\,s^{-1}}}\right]^{\alpha}\,{\rm h^{-5/2}_{70}\,Mpc^2}
\label{eq:ylx}
\end{equation}
The best fitting parameters are given in Table~\ref{tab:rellx}. The intrinsic scatter around the relation is important, more than $50\%$,  reflecting the important scatter,  at given $\YX$,  of the soft band luminosity computed without excising the core (see \citealt{pra09}).  
The best fitting relation is consistent with the relation expected from combining the $Y_{\rm sph}(\Rv)$--$\Mv$ relation derived from the universal pressure  profile (Eq.~\ref{eq:mysph})  and the  $\LX$--$\Mv$ relation. For consistency, the latter was updated (parameters given in Table~\ref{tab:rellx}), using  present $\Mv$  values derived from the updated \MY\ relation (Eq.~\ref{eq:mynstd}).   The slope and normalisation (taking into account the different pivot used)  are consistent with those published in \citet{pra09}.

%================================================================================
%================================================================================
\begin{table}[t]
     \caption[]{$h(z)^{-2/3}Y_{\rm sph}(\Rv)$--$h(z)^{-7/3}\LX$ and updated $h(z)^{-7/3}\LX$--$\Mv$ relations (see text). $\LX$ is the $[0.1$--$2.4]\,\keV$ luminosity  within $\Rv$. MB: relations corrected for Malmquist bias. For each observable set, $(B,A)$, we fitted a power law relation of the form $B  = C(A/A_0)^\alpha$, with $A_0 = 10^{44}\,{\rm h_{70}^{-2}}$\,ergs/s and $3\times10^{14}\,{\rm h_{70}^{-1}}\,\msol$ for $\LX$ and
 $\Mv$, respectively.   $\sigma_{\rm log,i}$: intrinsic scatter about  the best fitting relation in the $\log$--$\log$ plane.} 
     \label{tab:rellx}
     \begin{center}
    \begin{tabular}{lcccc}
    \hline
    \hline
    Relation &  $\log_{10} C $ & $\alpha$ & $\sigma_{\rm log,i}$ \\% & $\chi^2$\\
    \hline
$Y_{\rm sph}(\Rv)$--$\LX$--MB 	&  $-4.940 \pm 0.036$ &$1.07 \pm 0.08$ & $0.190\pm0.025$ \\ 
$\LX$--$\Mv$--MB  					   	&  $0.193 \pm 0.034$ &$1.76 \pm 0.13$ & $0.199\pm0.035$ \\ 
$Y_{\rm sph}(\Rv)$--$\LX$ 		&  $-5.047 \pm 0.037$ &$1.14 \pm 0.08$ & $0.184\pm0.024$ \\
$\LX$--$\Mv$ 			   				& $0.274 \pm 0.032$ &$1.64 \pm 0.12$ & $0.183\pm0.032$ \\      
\hline
    \end{tabular}
 \end{center}
\end{table}
%================================================================================
%================================================================================

For practical purposes, the scaling of  $\YSZ(x\,\Rv)\,D_{\rm A}^2$ or that of the total SZ signal with $\LX$ is of more direct interest than the $Y_{\rm sph}(\Rv)$--$\LX$ relation. In view of the good agreement of the latter with the universal profile model,  the $Y$--$\LX$ relation, for any integration region  of  interest, can be safely derived by correcting the normalisation in  Eq.~\ref{eq:ylx} by the model ratio of $Y$ to $Y_{\rm sph}(\Rv)$.  This ratio is simply $I(x)/I(1)$ for the spherically integrated Compton parameter, e.g., $I(5)/I(1)=1.796$ for the total SZ signal,   and $J(x)/I(1)$ for the $\YSZ(x\,\Rv)\,D_{\rm A}^2$ signal. 

\subsection{Comparison with standard self-similar relations}

The $Y$-$\Mv$ relations derived above do not seem to deviate much from standard self-similarity (Fig.~\ref{fig:ym}). A fully consistent standard (ST) model, with standard slope $Y$-$\Mv$ relations, is obtained when using the standard slope \MY\ relation (Eq.~\ref{eq:mystd}), as shown in Appendix~\ref{ap:std}.  The  universal profile and scaling relations obtained in that case are given in the Appendix, together with a detailed comparison of the presently derived scaling relations with the ST relations. In summary, the difference for the $Y$-$\Mv$ relations mirrors  that  for the  \MY\  relation. As compared to values derived  from the ST relation, $Y$ is lower at low mass and higher at high mass.   Typically, the difference for the total $Y_{SZ}$ signal ranges from   $-19\%$ to $+6\%$ in the $[10^{14}$--$10^{15}]\,\msol$ mass range.  On the other hand,
 the $Y$--$\LX$ relations, which  only depend on cluster internal structure,  are  essentially the same in the two models : the difference is less than $5\%$ in the $[0.1$--$10] 10^{44}$ ergs/s luminosity range.  

%-----------------------------------------------------------------------------------------------------------------------------------------------------------------------------------
%-----------------------------------------------------------------------------------------------------------------------------------------------------------------------------------
\section{Discussion and conclusions} 

The present work is the first examination of the properties of the ICM pressure for a representative sample of nearby clusters. The sample, \rexcess, was chosen by X-ray luminosity alone, without regard to morphology or dynamical state. It  covers the mass range $10^{14} < M_{500} < 10^{15}$ M$_\odot$, with mass iteratively estimated from the \MY\ relation, calibrated from a sample of relaxed clusters including \rexcess\ objects.  As for the entropy \citep{pra09b}, the depth of the observations allowed us to probe the scaling behavior of the pressure profiles out to $\Rv$. This is essential for a complete picture of the modification of the standard self-similarity due to non-gravitational processes, including its radial behavior. \\

Scaling the individual pressure profiles by mass and redshift according to the standard self-similar model, we derived an average scaled pressure profile for the cluster population and  relate the deviations about the mean to both the mass and the thermo-dynamical state of the cluster:
\begin{itemize}
\item Cool core systems exhibit more peaked profiles, while morphologically disturbed systems have shallower profiles.
\item As a result, the dispersion is large in the core region, reaching approximately 80 per cent at $0.03\,\Rv$.  However, as compared to the density, the pressure exhibits less scatter, a result of the anticorrelation of the density and temperature profiles interior to $0.2\,\Rv$. Outside the core regions, the dispersion about the average profile is remarkably low, at less than 30 per cent beyond $0.2\,\Rv$. 
\item We find  a residual mass dependence of the scaled profiles, with a slope of $\sim 0.12$, consistent with that expected from the empirical non-standard slope of the \MY\ relation. However, there is some evidence that  the departure from standard scaling decreases with radius and is consistent with zero at $R_{500}$.  We provide an analytical correction to the mean slope that accounts for this second order effect. 
\end{itemize}
The behaviour of the pressure profiles,  with respect to  standard self-similarity with zero dispersion, resembles that generally found for other quantities such as the entropy or density: 1) regularity in shape outside the core 2) increased dispersion inside the core linked to cooling effects and dynamical state and 3) departure from standard mass scaling that becomes less pronounced towards the cluster outskirts. However, the latter two  deviations are less pronounced than for the entropy and/or density, showing that the pressure is the quantity least affected by dynamical history and  non-gravitational physics. This further supports the view that  $Y_{\rm SZ}$ is indeed a  good mass-proxy. 

 Furthermore, our direct measure of the  $Y_{\rm sph}(\Rv)$--$\YX$ relation, where  $Y_{\rm sph}(\Rv)$ is the spherically integrated pressure profile,  exhibits  dispersion consistent with the $<5\%$ statistical scatter.  This shows that  variations in pressure profile shape do not introduce significant extra intrinsic scatter into the  $Y_{\rm sph}(\Rv)$--$\Mv$ relation as compared to that from the $\YX$--$\Mv$ relation. \\

The observational data are compared to and combined  with simulated data to derive the universal ICM pressure profile. This profile is then used to predict  the scaling relations involving the integrated Compton parameter $Y$. We consider both the spherically integrated quantity, $Y_{\rm sph}(R)$, which is related to the gas thermal energy, and also the cylindrically integrated quantity $Y_{\rm cyl}(R) = Y_{SZ}(R) D_A^2$, which is directly related to the observationally-derived SZ signal within $\theta=R/D_{\rm A}$:
\begin{itemize}
\item Simulated scaled profiles from three independent sets of state of the art numerical simulations show excellent agreement, within $20\%$, between $0.1$ and $3\Rv$, for pressures varying by 4 orders of magnitude in that radial range. 
\item Comparison with observed scaled data shows good agreement outside the core regions, which is the most relevant aspect for the $Y_{\rm SZ}$ estimate. The average simulation profile lies parallel to the observed data, with only a slight offset ($\sim 10$ per cent)  when the simulated profiles are scaled using the hydrostatic mass.
\item This motivates us to combine the average observed  scaled profile in the $[0.03-1]\,\Rv$ radial range with the average simulated profile  in the $[1-4]\,\Rv$ range. This hybrid profile is fitted by a generalised NFW model, which allows us to define a dimensionless universal ICM pressure profile. Combined with  the empirical mass scaling of the profiles, this universal profile defines the physical pressure profile of clusters, up to the cluster boundary, as a function of mass and redshift, assuming self-similar evolution.
\item  This universal profile allows us to derive the expected $Y_{\rm sph}(x\Rv)$--$\Mv$  or $\YSZ(x\Rv)$--$\Mv$  relations for any aperture. The 
slope is the inverse of the empirical slope of the \MY\ relation. The normalisation is given by the dimensionless integral of the universal profile within the region of interest expressed in scaled radius.  The corresponding  \YLX\  relations can be derived by combining the relevant $Y$--$\Mv$ relation with the empirical $\LX$--$\Mv$ relation. 
\item  The $Y_{\rm sph} (\Rv)$--$ \Mv$ and $Y_{\rm sph} (R_{500})$-- $\LX$ relations derived directly from the individual profiles are in excellent  agreement with those expected from the universal profile.
\item We confirm that the isothermal $\beta$--model over-estimates the $Y$ signal at given mass. This overestimate depends strongly on the assumption on cluster extent and reaches a factor of nearly two at $2\Rv$.
\end{itemize}
The convergence  of  various approaches to determine scaled  cluster profiles supports the robustness of our determination of the  universal pressure profile, particularly its shape. This includes the agreement  between independent simulations, between these simulations and the present observed data based on a representative cluster sample, and also the agreement between the  present \xmm\ data and published  \chandra\ data for clusters of similar thermo-dynamical state. As a result, we believe that  quantities which purely depend on the universal profile shape are particularly robust and well converged. This includes the typical SZ decrement  profile or relations between the Compton parameter  estimated in various apertures. 
 
However, the pressure profile interior to  $\Rv$ is derived from temperatures estimated using azimuthally averaged spectra. These have been corrected for the spectroscopic bias due to projection but not for azimuthal variations. In the cluster outskirts,  the electron-proton 
equilibration time is larger than the Hubble time \citep{fox97} and if the electron temperature is indeed smaller than the ion temperature, this will affect the pressure profile and  lead to a decrease in the total $Y_{SZ}$ signal \citep{rud09}.  
High resolution SZ data  with improved sensitivity are needed to probe any remaining systematic effects due to the spectroscopic bias, and to directly observe the shape of the pressure profile beyond $\Rv$, which is out of reach of current X--ray observatories.

Using the universal profile, the absolute normalisation and slope of the $Y$--$\Mv$ relations rely on  the underlying  observationally defined  \MY\ relation.  Initial comparison with $Y_{\rm SZ}(\Rv)$ data for 3 high mass systems,  measured with SZA by \citet{mro09} and  analysed with a realistic analytic pressure profile, indicates good agreement.  A key point is to extend this type of analysis to larger samples and  to include lower mass systems.  We further emphasize that the  \MY\ relation  was calibrated from  hydrostatic mass estimates  using relaxed objects. The $Y$--$\Mv$ relation we derive is  technically a $Y$--X--ray mass relation and  is expected to differ from the `true' $Y$--$\Mv$ by the offset between the `true' mass and the  hydrostatic mass for {\it relaxed} objects. 

A major open issue is the pressure evolution.  With the present study based on a local cluster sample, we could only assume standard self-similar evolution.  Because the SZ signal is not subject to redshift dimming, on going SZ surveys are expected to detect many new clusters at high z. Of particular interest is the  \planck\ survey, which, thanks to its All--Sky coverage, will detect massive, thus rare, clusters, the best objects for precise cosmology with clusters.  SZ follow-up, at the best possible resolution, and sensitive X--ray follow-up (particularly with \xmm) will be crucial to assess  possible evolution of pressure profile shape and  measure the evolution of the \MY\ and $Y_{SZ}$--$\Mv$ relations. \\

As a matter of practical application,  the universal pressure profile is given in Eq.~\ref{eq:pgnfw} with parameters in Eq.~\ref{eq:pargnfw}. For clusters of given mass $\Mv$ and $z$, the physical pressure profile can then be derived from Eq.~\ref{eq:puniv} and the spherical $Y_{\rm sph} (R)$ or  cylindrical $Y_{\rm cyl} (R)$ quantities can be estimated for any radius of interest using Eq.~\ref{eq:mysph}--\ref{eq:Ix} and Eq.~\ref{eq:mysz}--\ref{eq:Jx}, respectively.  

These equations can be used as is when  $\Mv$ is estimated for relaxed systems using the HSE equation,  and for all clusters using $\Mv$ derived from mass-proxy  relations. The preferential relations would be the  \MY\ and  the  $\Mv$--$\LX$, where $\LX$ is the core--excised bolometric luminosity \citep{pra09}, as both these relations display low scatter, compared to the relation between $\Mv$ and the full aperture soft band $\LX$.  A typical application would be to predict the SZ signal of a known X--ray cluster with measured $\LX$ or $\Mv$, or to estimate the mass and thus X-ray properties of newly discovered SZ clusters. Other applications include the analysis of low S/N and/or poor resolution  SZ observation of X-ray clusters, e.g., allowing to optimise  the integration aperture and use a realistic decrement shape. 

On the other hand, care is needed when knowledge of the 'true' mass is important, e.g., in predicting cluster number counts for future SZ surveys or  in SZ selection function modelling. The above total $Y_{\rm SZ}$--$\Mv$ relation should be corrected by the bias between the true mass and the HSE mass at $\Rv$, which is typically $\sim 13\%$ as determined from comparison with current numerical simulations.  Further progress on this fundamental question, as well as on the intrinsic scatter of the $Y$--$M$ relation,  is expected from the wealth of high quality multi-wavelength data that will be available in the coming years. 

\begin{acknowledgements}
We would like to thank Stefano Borgani, Daisuke Nagai, 
and Riccardo Valdarnini  for providing us with the simulations data 
and for helpful discussions  and useful comments on the manuscript.
The present work is based on observations obtained with {\it XMM-Newton}, an ESA science mission with instruments and contributions directly funded by ESA Member States and the USA (NASA).
EP acknowledge the support of grant ANR-06-JCJC-0141.
\end{acknowledgements}

%-----------------------------------------------------------------------------------------------------------------------------------------------------------------------------------
%-----------------------------------------------------------------------------------------------------------------------------------------------------------------------------------
\begin{appendix} 
\section{Characteristic self-similar quantities}
\label{ap:ss}

Following \citet{nag07} and \citet{voi05} the characteristic quantities, $P_{500}$ and $Y_{500}$, used in the present work, are defined from a simple self-similar model.  
The characteristic temperature is ${\rm k}T_{500}= \mu\,m_{\rm p}\,G\,\Mv/2\,\Rv$, the temperature 
of a singular isothermal sphere with mass $\Mv$. Here, $\mu$ is the mean molecular weight  and $m_{\rm p}$, the proton mass. We recall that $ \Mv$ is defined as the mass  within the radius $\Rv$ at which the
mean mass density is 500 times the critical density, $\rho_{\rm c}(z)$, of the universe at the
cluster redshift: $\Mv =  (4\pi/3)\,\Rv^3\,500\,\rho_{\rm c}(z)$ with  $\rho_{\rm c}(z)= 3 H(z)^2/ (8\pi\,G)$. $H(z)$ is the Hubble constant, $H(z) = H(0)\sqrt{\Omega_{\rm M}\,(1+z)^3 +\Omega_{\Lambda}}$ and $G$ is the Newtonian constant of gravitation. The characteristic gas density is  $\rho_{\rm g,500}= 500\,f_{\rm B}\,\rho_{\rm c}(z)$, i.e., the ratio of the gas density to the dark matter density is that of the Universe baryon fraction $f_{\rm B}$.  The electron density is $n_{\rm e,500} = \rho_{g,500}/(\mu_{\rm e}\,m_{\rm p})$ where $\mu_{\rm e}$  is the mean molecular weight  per free electron.  

\noindent The characteristic pressure, $P_{500}$, is then defined as: 
\begin{eqnarray}
P_{500} & =  & n_{\rm e,500}\,{\rm k}T_{500} \\
	& = & \frac{3}{8\pi}\,\left[\frac{500\,{\rm G^{-1/4}}\,{\rm H(z)^2} }{2}\right]^{4/3}\,\frac{\mu}{\mu_{\rm e}}\,f_{\rm B}\,\Mv^{2/3}
\end{eqnarray}
and  the corresponding characteristic integrated Compton parameter is:
\begin{eqnarray}
Y_{500} &  = &  \frac{\sigma_{\rm T}}{m_{\rm e}\,c^2}\,\frac{4\pi}{3}\,\Rv^3\,P_{500}  =  \frac{\sigma_{\rm T}}{m_{\rm e}\,c^2}\,\frac{ f_{\rm B}\,\Mv\,{\rm k}T_{500}}{\mu_{\rm e}\,m_{\rm p}}\\
 &  = & \frac{\sigma_{\rm T}}{m_{\rm e}\,c^2}\,\left[\frac{\sqrt{500}\,{\rm G}\,{\rm H(z)}}{4}\right]^{2/3} \,\frac{\mu}{\mu_{\rm e}}\,f_{\rm B}\,\Mv^{5/3} 
\end{eqnarray}

Numerical coefficients given in the corresponding Eq.~\ref{eq:p500} and \ref{eq:y500} are obtained for $f_{\rm B}=0.175$, $\mu= 0.59$ and $\mu_{\rm e}=1.14$, the values  adopted by \citet{nag07}, allowing a direct comparison with their best fitting GNFW model.  Note that the exact choice for these parameters does not matter, and does not need to reflect 'true' values, as long as the same convention is used throughout the study (e.g., when comparing observed and theoretical scaled profiles or observed scaled profiles from different samples or instruments).  

%-----------------------------------------------------------------------------------------------------------------------------------------------------------------------------------
%-----------------------------------------------------------------------------------------------------------------------------------------------------------------------------------
\section{The standard self-similar  case}
\label{ap:std}

In this Appendix, we summarise results  (hereafter ST results) obtained when $\Mv$ is estimated for  each \rexcess\ clusters using the \MY\ relation with a standard slope  (Eq.~\ref{eq:mystd}). The other physical  parameters are consistently estimated, $\Rv$, $\YX$ and  $\TX$  simultaneously in the iteration process used to derive $\Mv$ (Sec.~\ref{sec:data}),  and $Y_{\rm sph}(\Rv)$  from integration of the pressure profiles up to $\Rv$. For practical purposes, the baseline parameters  obtained using the best fitting empirical \MY\ relation (Eq.~\ref{eq:mynstd}) can be converted to the ST values  using  the power law relations given in Table~\ref{tab:stdnstd}. The luminosity  $\LX$ is kept unchanged, the difference in $\Rv$ values (at most $4.5\%$), having  a negligible impact due to the steep drop of emission with  radius.

%================================================================================
%================================================================================
\begin{figure}[t]
\centering
\includegraphics[width=\columnwidth, keepaspectratio]{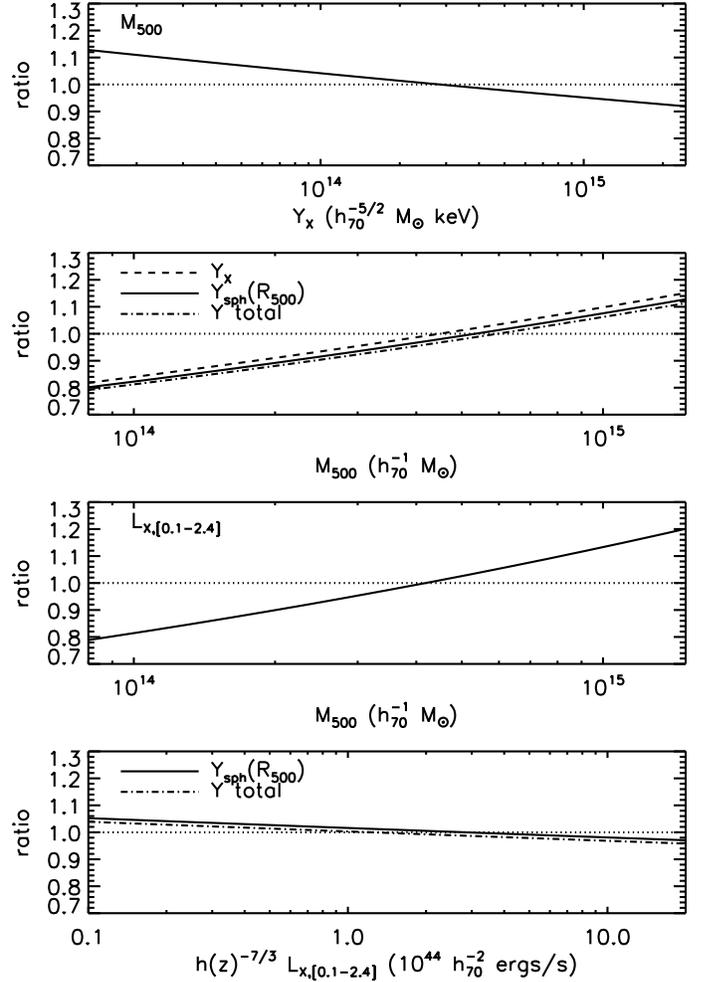} 
 \caption{\footnotesize Ratio of the scaling relations derived  using the empirical \MY\ relation (Eq.~\ref{eq:mynstd}) to those derived using the standard slope relation (Eq.~\ref{eq:mystd}). From top to bottom: $\Mv$ as a function of $\YX$;   $\YX$, $Y_{\rm sph}(\Rv)$ and total $Y_{\rm SZ}$  as a function of $\Mv$; $\LX$ as a function of  $\Mv$;  $Y_{\rm sph}(\Rv)$ and total $Y_{\rm SZ}$ as a function of $\LX$.} 
 \label{fig:comp}
\end{figure}
%================================================================================
%================================================================================

%================================================================================
%================================================================================
\begin{table}[b]
     \caption[]{ Power law relations to convert physical parameters of  \rexcess\ clusters from those  derived using the empirical \MY\ relation (Eq.~\ref{eq:mynstd}) to those derived using the standard slope relation (Eq.~\ref{eq:mystd}). For each observable, $Q$, the conversion follows the form $Q^{\rm ST}  = C(Q/Q_0)^\alpha$ where the pivot, $Q_{0}$ is  $3\times10^{14}\msol$, $5\,\keV$,    $2\times10^{14}\msol\,\keV$ and $2\times10^{-5}\,{\rm Mpc^2}$ for $\Mv$,$\TX$, $\YX$ and  $Y_{\rm sph}(\Rv)$, respectively. } 
     \label{tab:stdnstd}
     \begin{center}
    \begin{tabular}{lcc}
    \hline
    \hline
    Relation &  $C $ & $\alpha$ \\
    \hline
$\Mv^{\rm ST}$--	$\Mv$&  $0.968$ &$1.089$  \\ 
$\TX^{\rm ST}$--$\TX$ 	&  $1.002$ &$0.992$  \\ 
$\YX^{\rm ST}$--$\YX$	& $0.995$ &$1.017$  \\ 
$Y_{\rm sph}^{\rm ST}(\Rv)$--$Y_{\rm sph}(\Rv)$	& $0.991$ &$1.031$  \\ 
\hline
    \end{tabular}
 \end{center}
\end{table}
%================================================================================
%================================================================================

In the ST case, the scaled pressure profiles do not show any significant dependence on mass,  as shown in Sec.~\ref{sec:pscmass}.
In other words, the pressure profiles follow a standard self-similar mass scaling:
\begin{equation}
\label{eq:punivstd}
P(r)  =   P_{500} \,\mathbb{p}\left(r/\Rv\right)
\end{equation}
with $P_{500}$ defined by Eq.~\ref{eq:p500}. The GNFW parameters of the universal profile $\mathbb{p}(x)$, derived as described in Sec.~\ref{sec:puniv}, are:

{\small
\begin{equation}
[P_{0} ,c_{500},\gamma,\alpha,\beta] =  [8.130\,{\rm h_{70}^{-3/2}}, 1.156,0.3292,1.0620,5.4807]
\label{eq:pargnfwstd}
\end{equation}
}

As a result, the integrated Compton parameters also follow standard self-similarity, $Y\propto\Mv^{5/3}$. The $Y$--$\Mv$ relations derived from the universal pressure profile can be written as: 
\begin{eqnarray}
\label{eq:myszstd}
Y_{\rm sph}(x\,\Rv) &  =  &Y_{500}\,I(x) \\ 
 \YSZ(x\,\Rv)\,D_{\rm A}^2  &  =  &Y_{500}\,J(x) \nonumber
\end{eqnarray}
with  $Y_{500}$ given by Eq.~\ref{eq:y500} and $I(x)$ or $J(x)$ defined by Eq.~\ref{eq:Ix} and Eq.~\ref{eq:Jx}, respectively. For the GNFW parameters given by Eq.~\ref{eq:pargnfwstd},  the numerical values of $I(1), I(5)$ and $J(1)$ are $0.6552$, $1.1885$ and $0.7913$, respectively.  The  $Y_{\rm sph}(x\,\Rv)$--$\Mv$ relation derived from a direct fit to the data has a slope of  $1.663\pm0.013$, fully consistent with $5/3$. Over the $[10^{14}$--$10^{15}]\,\msol$ mass range, it differs by less than $0.8\%$ from that derived from the universal profile (Eq.~\ref{eq:myszstd}).

%================================================================================
%================================================================================
\begin{table}[t]
     \caption[]{$h(z)^{-2/3}Y_{\rm sph}(\Rv)$--$h(z)^{-7/3}\LX$ and $h(z)^{-7/3}\LX$--$\Mv$ relations for $\Mv$ estimated using the standard slope \MY\ relation (Eq.~\ref{eq:mystd}).  Same notations as in Table~\ref{tab:rellx}.} 
     \label{tab:rellxstd}
     \begin{center}
    \begin{tabular}{lcccc}
    \hline
    \hline
    Relation &  $\log_{10} C $ & $\alpha$ & $\sigma_{\rm log,i}$ \\% & $\chi^2$\\
    \hline
$Y_{\rm sph}(\Rv)$--$\LX$--MB 	&  $-4.947 \pm 0.037$ &$1.08 \pm 0.08$ & $0.192\pm0.025$ \\ 
$\LX$--$\Mv$--MB  					   	&  $0.215 \pm 0.035$ &$1.61 \pm 0.12$ & $0.199\pm0.035$ \\ 
$Y_{\rm sph}(\Rv)$--$\LX$ 		&  $-5.056 \pm 0.038$ &$1.16 \pm 0.08$ & $0.184\pm0.024$ \\
$\LX$--$\Mv$ 			   				& $0.295 \pm 0.032$ &$1.50 \pm 0.11$ & $0.183\pm0.032$ \\      
\hline
    \end{tabular}
 \end{center}
\end{table}
%================================================================================
%================================================================================

We also derived the observed $Y_{\rm sph}(\Rv)$--$\LX$ relation, as well as the $\LX$--$\Mv$ corresponding to the modified $\Mv$ values. The best fitting power law parameters are given in Table~\ref{tab:rellxstd}. The former is  consistent with the relation expected from combining the $\LX$--$\Mv$ relation with the $Y_{\rm sph}(\Rv)$--$\Mv$ relation derived from the universal pressure  profile (Eq.~\ref{eq:myszstd}).  The $Y$--$\LX$ relation, for any integration region  of  interest, can be derived by correcting the normalisation of the $Y_{\rm sph}(x\,\Rv)$--$\LX$ given in Table~\ref{tab:rellxstd} by the model ratio  of $Y$ to $Y_{\rm sph}(\Rv)$, as described in Sec.~\ref{sec:ylx}.

Figure~\ref{fig:comp} compares the scaling relations derived in the paper with the ST relations derived in this section.  The empirical slope of the \MY\ relation being smaller than the standard value, $\Mv$ at a given $\YX$ is higher at low $\YX$ and smaller at high $\YX$ (top panel). Equivalently, $\YX$ at given mass is smaller at low mass, by $\sim -16\%$ at $\Mv=10^{14}\msol$, 
and higher at high mass, by $\sim +10\%$ at $\Mv=10^{15}\msol$ 
(second panel). The behavior of $Y_{\rm SZ}$ closely follows that of $\YX$ (same panel) simply because the ratio of the two purely depends on the shape of the universal profile. This shape is  barely affected by the small difference in $\Rv$ values used to scale the physical pressure profiles.
Similarly, the $Y_{\rm SZ}$--$\LX$ relation only depends on cluster internal structure and is essentially the same in the two models (bottom panel).   $Y_{\rm SZ}(\Rv)$ is slightly higher/lower at low/high $\LX$ following the change of  $\Rv$ at given $\LX$. As the  \MY\  is shallower than the ST relation,  the  $\Mv$--$\LX$  is also shallower (thus higher $\Rv$  at low $\LX$) or equivalently the $\LX$--$\Mv$  is steeper (third panel). 

%-----------------------------------------------------------------------------------------------------------------------------------------------------------------------------------
%-----------------------------------------------------------------------------------------------------------------------------------------------------------------------------------
\section{Pressure profiles and best fitting model }

Here we list the physical cluster  properties and the parameters of the GNFW  model best fitting each profile (Table~\ref{ap:pnfw}). 
Individual profiles and their best fitting model are plotted in Fig.~\ref{fig:pfita}--\ref{fig:pfitc}.
\label{ap:pnfw}
\begin{table*}[b]
     \caption[]{Cluster physical parameters.  Column (2)-(3): $\Rv$ is the radius corresponding to a density contrast of 500,  estimated iteratively from the \MY\ relation (Eq.~\ref{eq:mynstd}), where $Y_{\rm X}=\Mgv\TX$ is the product of the gas mass within $\Rv$ and the spectroscopic temperature $\TX$.   Column (3) and (4): spherically integrated Compton parameter within $R_{2500}$ and $\Rv$, respectively. Column (5):  $P_{500}$ as defined by Eq.~\ref{eq:p500}. Column (6) to (9)  give the best fitting GNFW parameters for the pressure profiles (Eq.~\ref{eq:pgnfw}). The external slope parameter $\beta$ has been fixed to $5.49$ (see text). Redshift $z$ and $\Mv$ values can be found in Table~1 of \citet{pra09b}. }
     \label{tab:pnfw}
     \centering
    \begin{tabular}{lccccccccccc}
    \hline
    \hline
    Cluster &  $\Rv $ & $Y_{\rm X}$ & $Y_{\rm sph}(R_{2500})$ &  $Y_{\rm sph}(\Rv)$ & $P_{500}$ & $P_0$&$c_{500}$ & $\alpha$ & $\gamma$ &$\chi^2/{\rm dof}$\\
     & (Mpc) & $(10^{14}\,\msol\,\keV)$ & $(10^{-5}\,{\rm Mpc^2})$ &$(10^{-5}\,{\rm Mpc^2})$& $(10^{-3}\,\keV\,{\rm cm}^{-3})$ & \\
    \hline
RXC J0003.8+0203 & $ 0.879$ & $ 0.763\pm0.030$ & $ 0.410\pm0.009$ & $ 0.990\pm0.036$ & $ 1.466$ & $ 3.93$ & $1.33$ & $1.41$ & $0.567 $ & $0.3/9$ \\
RXC J0006.6-3443 & $ 1.075$ & $ 2.35\pm0.13$ & $ 1.030\pm0.050$ & $ 3.06\pm0.16$ & $ 2.292$ & $ 3.27$ & $1.10$ & $1.41$ & $0.408 $ & $0.0/1$ \\
RXC J0020.7-2542 & $ 1.056$ & $ 2.253\pm0.072$ & $ 1.419\pm0.034$ & $ 2.80\pm0.11$ & $ 2.331$ & $ 20.26$ & $2.16$ & $1.37$ & $0.035 $ & $3.7/7$ \\
RXC J0049.4-2931 & $ 0.800$ & $ 0.477\pm0.022$ & $ 0.277\pm0.010$ & $ 0.630\pm0.037$ & $ 1.254$ & $ 8.58$ & $1.31$ & $1.07$ & $0.422 $ & $0.2/4$ \\
RXC J0145.0-5300 & $ 1.112$ & $ 2.819\pm0.097$ & $ 1.193\pm0.029$ & $ 3.89\pm0.18$ & $ 2.461$ & $ 9.73$ & $1.06$ & $1.06$ & $0.000 $ & $1.1/4$ \\
RXC J0211.4-4017 & $ 0.684$ & $ 0.203\pm0.006$ & $ 0.101\pm0.003$ & $ 0.267\pm0.010$ & $ 0.902$ & $ 8.97$ & $1.04$ & $0.93$ & $0.267 $ & $3.0/6$ \\
RXC J0225.1-2928 & $ 0.683$ & $ 0.185\pm0.014$ & $ 0.087\pm0.004$ & $ 0.237\pm0.017$ & $ 0.832$ & $ 19.28$ & $1.19$ & $0.88$ & $0.000 $ & $5.4/5$ \\
RXC J0345.7-4112 & $ 0.685$ & $ 0.188\pm0.009$ & $ 0.109\pm0.003$ & $ 0.227\pm0.009$ & $ 0.836$ & $ 3.68$ & $1.65$ & $1.67$ & $0.690 $ & $1.1/7$ \\
RXC J0547.6-3152 & $ 1.148$ & $ 3.59\pm0.11$ & $ 1.976\pm0.037$ & $ 4.54\pm0.14$ & $ 2.799$ & $ 8.52$ & $1.74$ & $1.51$ & $0.260 $ & $3.8/6$ \\
RXC J0605.8-3518 & $ 1.059$ & $ 2.285\pm0.070$ & $ 1.264\pm0.025$ & $ 3.13\pm0.14$ & $ 2.338$ & $ 4.23$ & $0.88$ & $0.96$ & $0.659 $ & $1.1/6$ \\
RXC J0616.8-4748 & $ 0.947$ & $ 1.194\pm0.044$ & $ 0.515\pm0.014$ & $ 1.627\pm0.060$ & $ 1.784$ & $ 4.06$ & $1.16$ & $1.43$ & $0.234 $ & $1.4/3$ \\
RXC J0645.4-5413 & $ 1.302$ & $ 7.291\pm0.248$ & $ 3.60\pm0.11$ & $ 9.93\pm0.47$ & $ 3.722$ & $ 11.10$ & $0.94$ & $0.89$ & $0.265 $ & $2.5/5$ \\
RXC J0821.8+0112 & $ 0.753$ & $ 0.325\pm0.017$ & $ 0.171\pm0.007$ & $ 0.400\pm0.019$ & $ 1.053$ & $ 1.72$ & $1.37$ & $2.01$ & $0.860 $ & $1.5/1$ \\
RXC J0958.3-1103 & $ 1.076$ & $ 2.64\pm0.25$ & $ 1.72\pm0.11$ & $ 3.42\pm0.40$ & $ 2.553$ & $ 4.13$ & $1.77$ & $2.07$ & $0.719 $ & $0.0/3$ \\
RXC J1044.5-0704 & $ 0.939$ & $ 1.189\pm0.024$ & $ 0.732\pm0.010$ & $ 1.550\pm0.051$ & $ 1.820$ & $ 7.08$ & $1.27$ & $1.05$ & $0.644 $ & $13.7/7$ \\
RXC J1141.4-1216 & $ 0.893$ & $ 0.879\pm0.018$ & $ 0.491\pm0.007$ & $ 1.199\pm0.046$ & $ 1.597$ & $ 4.42$ & $1.08$ & $1.08$ & $0.652 $ & $15.3/6$ \\
RXC J1236.7-3354 & $ 0.758$ & $ 0.335\pm0.011$ & $ 0.162\pm0.003$ & $ 0.479\pm0.020$ & $ 1.062$ & $ 47.76$ & $0.72$ & $0.61$ & $0.000 $ & $3.2/4$ \\
RXC J1302.8-0230 & $ 0.850$ & $ 0.625\pm0.020$ & $ 0.305\pm0.007$ & $ 0.800\pm0.039$ & $ 1.349$ & $ 3.63$ & $1.09$ & $1.21$ & $0.519 $ & $14.8/6$ \\
RXC J1311.4-0120 & $ 1.351$ & $ 9.27\pm0.17$ & $ 5.610\pm0.084$ & $ 11.60\pm0.30$ & $ 4.169$ & $ 23.13$ & $1.16$ & $0.78$ & $0.399 $ & $17.1/7$ \\
RXC J1516+0005 & $ 1.010$ & $ 1.689\pm0.050$ & $ 0.927\pm0.013$ & $ 2.211\pm0.083$ & $ 2.035$ & $ 4.48$ & $1.52$ & $1.65$ & $0.474 $ & $4.1/5$ \\
RXC J1516.5-0056 & $ 0.932$ & $ 1.105\pm0.038$ & $ 0.479\pm0.015$ & $ 1.494\pm0.054$ & $ 1.740$ & $ 2.57$ & $1.09$ & $1.51$ & $0.465 $ & $1.2/4$ \\
RXC J2014.8-2430 & $ 1.176$ & $ 4.133\pm0.097$ & $ 2.293\pm0.056$ & $ 5.59\pm0.23$ & $ 2.971$ & $ 4.94$ & $0.75$ & $0.82$ & $0.684 $ & $8.8/7$ \\
RXC J2023.0-2056 & $ 0.740$ & $ 0.281\pm0.014$ & $ 0.149\pm0.005$ & $ 0.358\pm0.016$ & $ 0.968$ & $ 4.00$ & $1.36$ & $1.41$ & $0.515 $ & $0.2/2$ \\
RXC J2048.1-1750 & $ 1.095$ & $ 2.782\pm0.084$ & $ 1.104\pm0.024$ & $ 3.73\pm0.12$ & $ 2.542$ & $ 4.34$ & $1.33$ & $1.76$ & $0.000 $ & $10.7/3$ \\
RXC J2129.8-5048 & $ 0.903$ & $ 0.856\pm0.043$ & $ 0.357\pm0.016$ & $ 1.147\pm0.051$ & $ 1.508$ & $ 9.21$ & $0.94$ & $1.00$ & $0.000 $ & $0.2/0$ \\
RXC J2149.1-3041 & $ 0.891$ & $ 0.864\pm0.024$ & $ 0.429\pm0.009$ & $ 1.135\pm0.051$ & $ 1.585$ & $ 9.96$ & $0.71$ & $0.71$ & $0.446 $ & $3.3/6$ \\
RXC J2157.4-0747 & $ 0.753$ & $ 0.311\pm0.012$ & $ 0.122\pm0.005$ & $ 0.411\pm0.015$ & $ 1.007$ & $ 1.46$ & $1.24$ & $2.54$ & $0.491 $ & $0.1/1$ \\
RXC J2217.7-3543 & $ 1.031$ & $ 2.023\pm0.050$ & $ 1.079\pm0.021$ & $ 2.611\pm0.077$ & $ 2.260$ & $ 27.70$ & $1.18$ & $0.81$ & $0.133 $ & $0.2/5$ \\
RXC J2218.6-3853 & $ 1.147$ & $ 3.51\pm0.14$ & $ 1.796\pm0.049$ & $ 4.94\pm0.29$ & $ 2.751$ & $ 27.29$ & $1.06$ & $0.82$ & $0.000 $ & $1.0/4$ \\
RXC J2234.5-3744 & $ 1.307$ & $ 7.22\pm0.17$ & $ 4.300\pm0.075$ & $ 8.82\pm0.25$ & $ 3.647$ & $ 25.04$ & $2.01$ & $1.23$ & $0.000 $ & $10.6/5$ \\
RXC J2319.6-7313 & $ 0.793$ & $ 0.445\pm0.018$ & $ 0.194\pm0.004$ & $ 0.612\pm0.026$ & $ 1.207$ & $ 338.9$ & $0.17$ & $0.33$ & $0.065 $ & $1.9/5$ \\
\hline                                   %inserts single line
\end{tabular}
\end{table*}

\onlfig{1}{
\begin{figure*}[b]
\centering
\includegraphics[ width=\hsize, keepaspectratio]{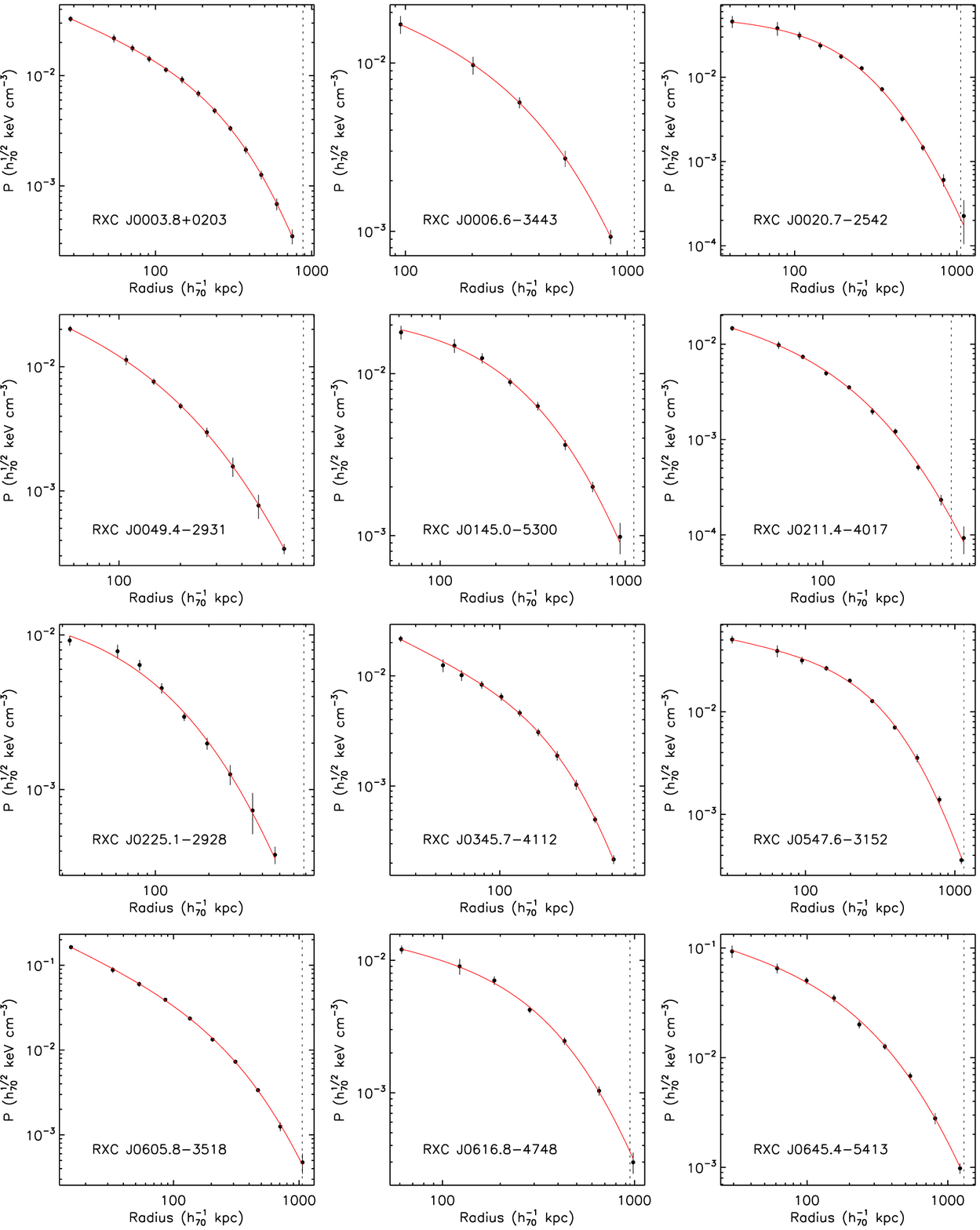} 
 \caption{\footnotesize Pressure profiles for the entire \rexcess\ sample with the best fitting GNFW model (red line). The dotted vertical line indicates $\Rv$ for each cluster.  } 
 \label{fig:pfita} 
\end{figure*}
}

\onlfig{2}{
\begin{figure*}[b]
\centering
\includegraphics[ width=\hsize, keepaspectratio]{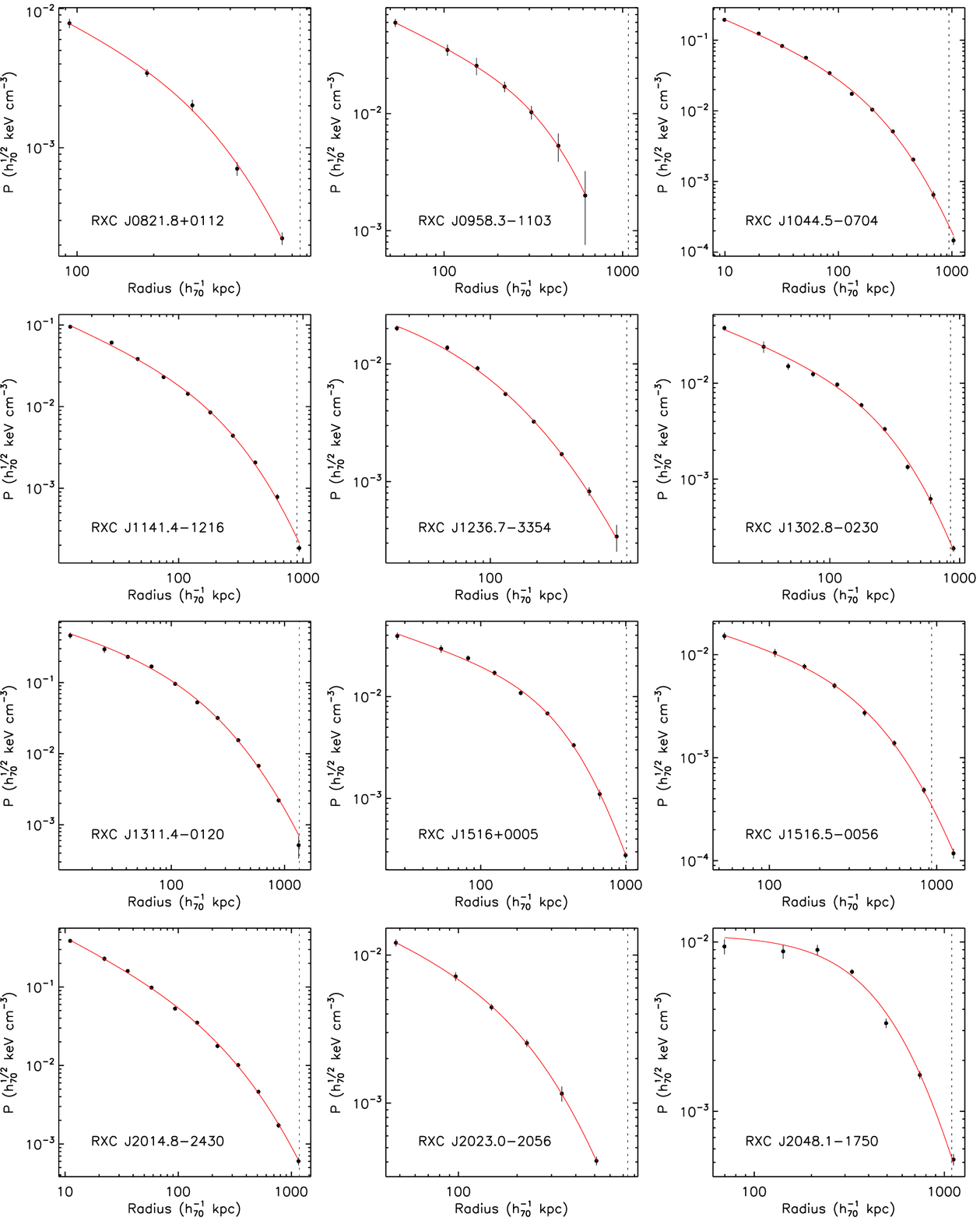} 
 \caption{\label{fig:pfitb}  \footnotesize continued } 
\end{figure*}
}

\onlfig{3}{
\begin{figure*}[b]
\centering
\includegraphics[ width=\hsize, keepaspectratio]{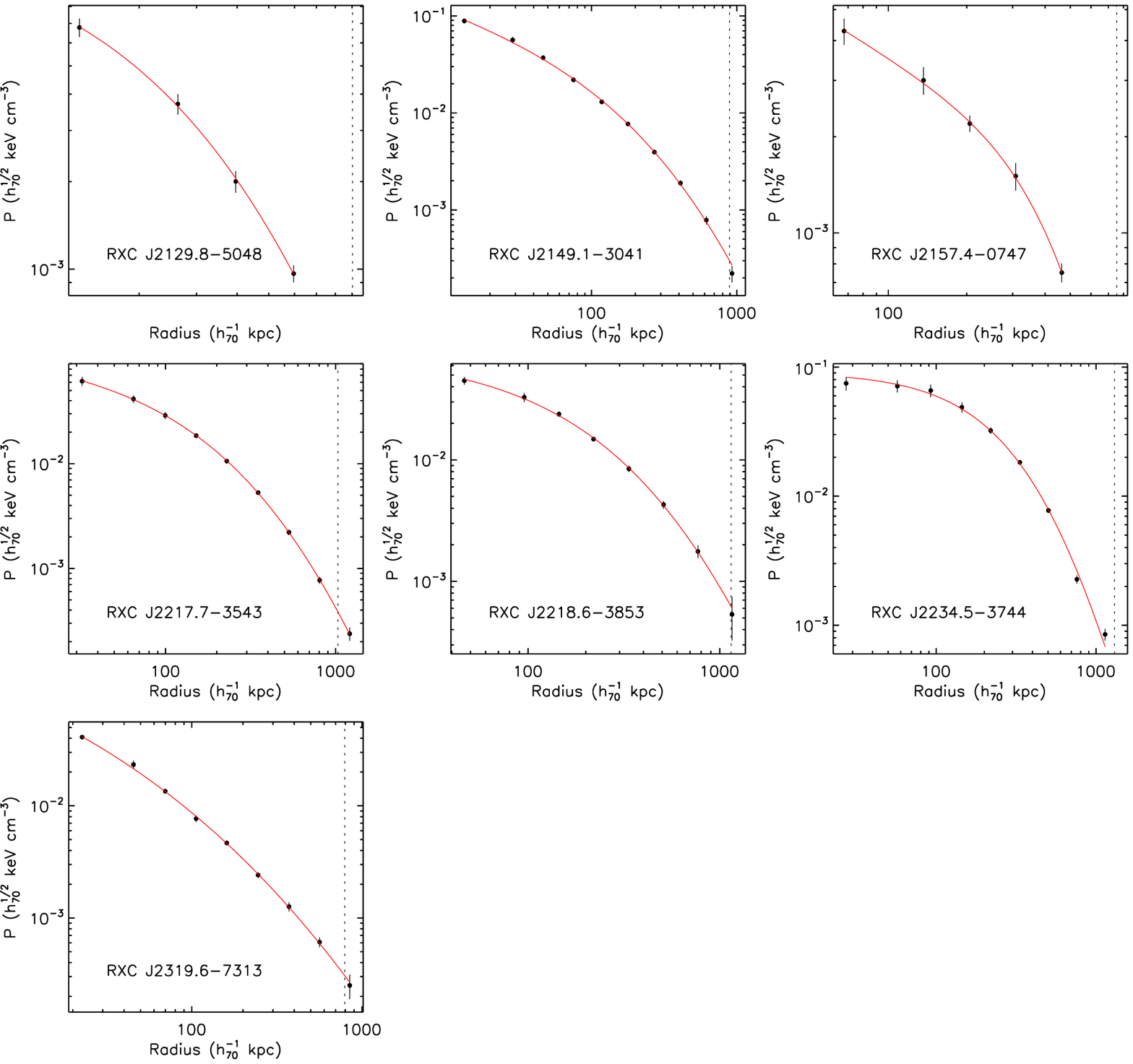} 
 \caption{\label{fig:pfitc}  \footnotesize continued } 
\end{figure*}
}

\end{appendix}

\end{document}